\documentclass[journal=jacsat,manuscript=article]{achemso}
\usepackage[version=3]{mhchem} 
\usepackage{blindtext}
\usepackage{pdfpages}
\usepackage{hyperref}
\usepackage{braket}
\usepackage{graphicx}
\usepackage{lineno}
\hypersetup{colorlinks=true, linkcolor=blue, citecolor=blue, urlcolor=blue}
\newcommand{\s}{WS$_2$}
\newcommand{\se}{WSe$_2$}

\newcommand{\colortext}{\textcolor{black}}

\usepackage{placeins}

\let\Oldsection\section
\renewcommand{\section}{\FloatBarrier\Oldsection}

\let\Oldsubsection\subsection
\renewcommand{\subsection}{\FloatBarrier\Oldsubsection}

\let\Oldsubsubsection\subsubsection
\renewcommand{\subsubsection}{\FloatBarrier\Oldsubsubsection}
\author{Anish Kumar}
\affiliation[Indian Institute of Science]
{Department of Electrical Communication Engineering, Indian Institute of Science, Bangalore, India, 560012}

\author{Suman Chatterjee}
\affiliation[Indian Institute of Science]
{Department of Electrical Communication Engineering, Indian Institute of Science, Bangalore, India, 560012}

\author{Kenji Watanabe}
\affiliation{Research Center for Electronic and Optical Materials, National Institute of Material Science, 1-1 Namiki, Tsukuba, 305-0044, Japan}
\author{Takashi Taniguchi}
\affiliation{Research Center for Materials Nanoarchitectonics, National Institute of Material Science, 1-1 Namiki, Tsukuba, 305-0044, Japan}

\author{Kausik Majumdar}
\email{kausikm@iisc.ac.in}
\affiliation[Indian Institute of Science]
{Department of Electrical Communication Engineering, Indian Institute of Science, Bangalore, India, 560012}

\title[]
  {Supermoir\'{e}-trapped quadrupolar exciton}
\begin{document}

\begin{abstract}
Moir\'e-trapped dipolar interlayer exciton in heterobilayers offers a rich platform to explore interaction-driven phenomena. Extending to three layers and beyond leads to highly intriguing multipolar exciton - a superposition state of vertically aligned phase-coherent excitons. However, in experiments, unintentional twist-angle mismatch among layers may degrade the strength and homogeneity of the vertical Coulomb coupling. Here we propose that the supermoir\'e effect in a heterotrilayer comes to the rescue by creating periodic pockets of vertically aligned atomic registries that facilitate the formation of quadrupolar excitons trapped in such pockets. Using \s/\se/\s {} stack, we show interaction between multiple confined levels of the top and bottom moir\'e interfaces, creating electric field tunable multi-level hybridized bright (symmetric) and dark (anti-symmetric) quadrupolar states. Our work underscores the critical role of supermoir\'e effect in quadrupolar excitons. The discovery of reduced sensitivity on precise angle-alignment will ignite exploration of complex excitonic states in multi-layered heterostructures. 
\\
\\
\textbf {KEYWORDS:} Quadrupolar exciton, supermoir\'e, moir\'e superlattice, trilayer, 2D materials.
\end{abstract}

\clearpage
Interactions play a central role in all physical systems, dictating the emergent properties - from microscopic to macroscopic - across all length scales. With their atomically thin nature, the two-dimensional semiconducting transition metal dichalcogenides (TMDs) and their twisted heterostructures have emerged as a state-of-the-art platform for investigating the emergent properties through tuning inter-particle interaction \cite{shahnazaryanExcitonexcitonInteractionTransitionmetal2017, erkenstenExcitonexcitonInteractionTransition2021, chenTuningMoireExcitons2022, chernikovElectricalTuningExciton2015, liContinuousMottTransition2021a, makSemiconductorMoireMaterials2022, julkuExcitonInteractingMoire2024, tranEvidenceMoireExcitons2019}. 
The optoelectronic properties of TMDs are governed by strongly bound excitons and their higher order complexes \cite{liRevealingBiexcitonTrionexciton2018, qiuOpticalSpectrumMoS2013}. These excitonic resonances are excellent probes to study interaction through optical readout \cite{mouriNonlinearPhotoluminescenceAtomically2014,liRevealingBiexcitonTrionexciton2018}. Stacking these TMDs to form a van der Waals heterostructure permits an additional degree of freedom, allowing for band engineering and layer-polarized interlayer excitons. Furthermore, the twist angle between the layers enables tunability of the interaction scale by forming a periodic moir\'e potential \cite{aldenStrainSolitonsTopological2013, heMoirePatterns2D2021, devakulMagicTwistedTransition2021a, zhangMoireQuantumChemistry2020}. For example, moir\'e confined excitonic complexes have been proven to be an excellent Hubbard model simulator in a rich phase space \cite{wuHubbardModelPhysics2018, slagleChargeTransferExcitations2020, tangSimulationHubbardModel2020, panBandTopologyHubbard2020}.
\par
\begin{figure}
\centering
\includegraphics[width = \textwidth]{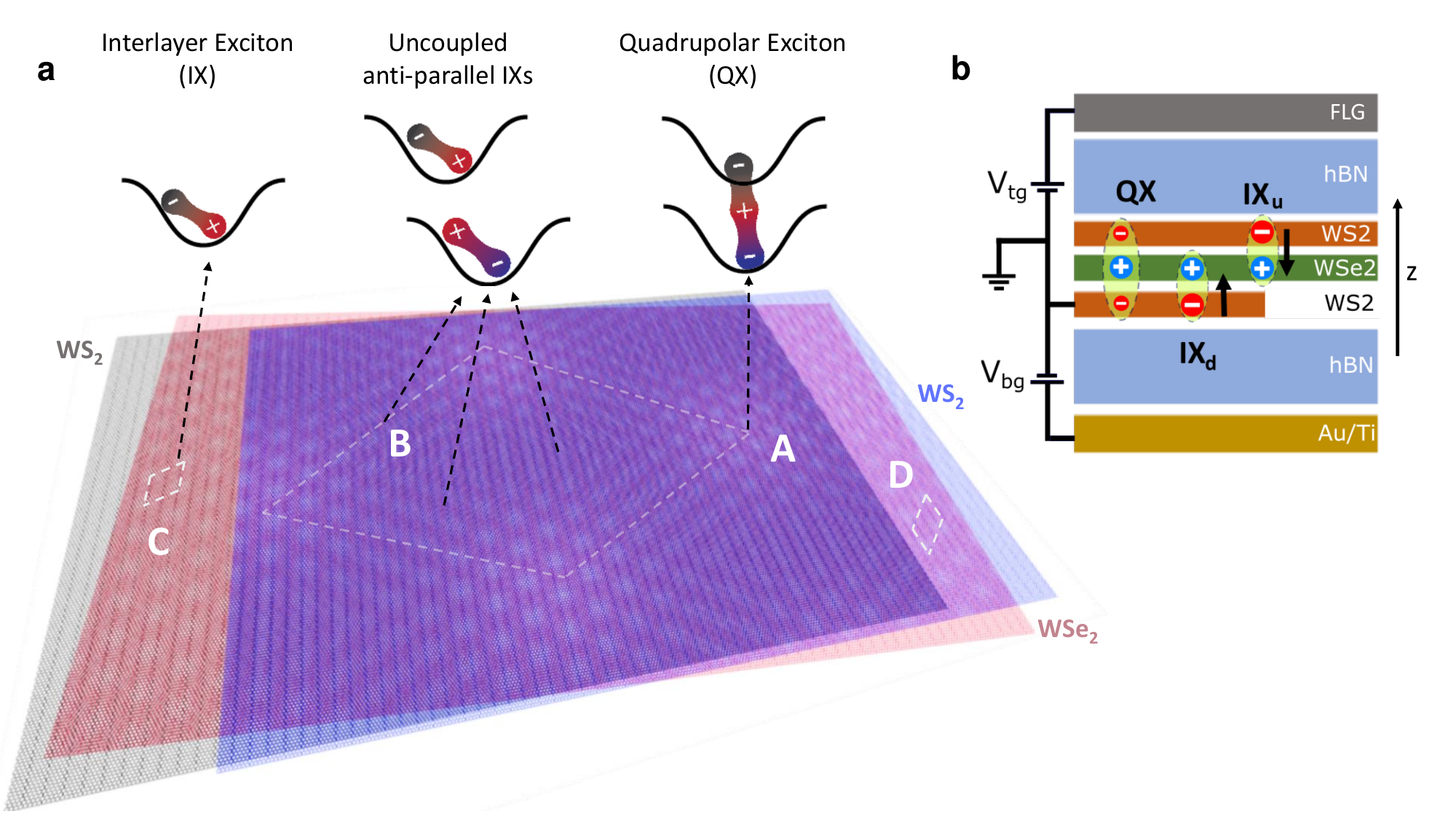}
\caption{\textbf{Supermoir\'e landscape in hetero-trilayer:}\textbf{(a)} Illustration of the moir\'e landscape in the heterostructure and different excitonic species existing at different local stackings. The bilayer regions (\s/\se) host interlayer excitons (IXs) at positions defined by the moir\'e, as shown by the white dashed lines (regions C and D). The trilayer region (\s/\se/\s) shows the emergence of a supermoir\'e and can host uncoupled IXs (in region B) and quadrupolar excitons (QXs) (in region A) at different positions. \textbf{(b)} Schematic of the dual gated heterostructure with top and bottom gate enabling independent doping and electric field control. The distribution of the electrons and holes for the quadrupolar exciton (QX), upper (IX$_u$) and lower (IX$_d$) interlayer excitons is shown. Smaller electrons in QX illustrate a single electron spread across the top and bottom layers. The black arrows represent the direction of the permanent out-of-plane dipole moments (which is zero for $QX$) of the corresponding exciton in the absence of an external electric field.}
\label{fig:1}
\end{figure}

Recently, an intriguing form of inter-layer exciton, called quadrupolar exciton, has been predicted \cite{slobodkinQuantumPhaseTransitions2020a, astrakharchikQuantumPhaseTransition2021a,  deilmannQuadrupolarDipolarExcitons2024} and experimentally realised \cite{liQuadrupolarDipolarExcitonic2023a, yuObservationQuadrupolarDipolar2023a, xieBrightDarkQuadrupolar2023, baiEvidenceExcitonCrystals2023, lianQuadrupolarExcitonsHybridized2023} in trilayer TMD heterostructures. An interlayer exciton in a typical heterobilayer (such as \s/\se~in \textbf{Fig. \ref{fig:1}}) exhibits a permanent out-of-plane dipole moment (electrons and holes being separated in two layers), while being polarizable by an in-plane oscillating field. Henceforth, we refer to these as interlayer dipolar exciton ($IX$). In contrast, a quadrupolar exciton ($QX$) in a nearly symmetric trilayer (such as \s/\se/\s~in \textbf{Fig. \ref{fig:1}}) is a superposition of two interlayer excitons having permanent anti-parallel dipole moments in the out-of-plane direction. This superposition of interlayer excitons manifests itself as a hyperbolic DC Stark effect in response to an out-of-plane electric field - a signature of hybridization. Interestingly, the $QX$ is still polarizable by an oscillating in-plane electric field, and hence can couple with normally incident light through dipolar interaction. Thus, a $QX$ can be thought of as an interlayer exciton with a zero permanent dipole moment in the out-of-plane direction and a non-zero in-plane transient dipole moment under oscillating field. This unique arrangement is predicted to host a multitude of exotic physics, such as rotonization and quadrupolar supersolid. \cite{astrakharchikQuantumPhaseTransition2021a}
\par
Note that, the electrons and holes in the individual layers reside on the Brillouin Zone corners. This requires the individual layers to be angle-aligned to each other to form vertically well-coupled excitonic states. Moreover, in the trilayer, two different moir\'e patterns form between the two interfaces, the top-middle and middle-bottom interface (see \textbf{Fig. \ref{fig:1}a}). The two moir\'e patterns interfere and produce a supermoir\'e pattern \cite{zhengLocalizationenhancedMoireExciton2023, wangCompositeSupermoireLattices2019a}. Given the requirement of closely angle-aligned layers and lattice constant mismatch between individual materials used, the formation of such moir\'e patterns is unavoidable in an experimental setup. Despite this, the effect of moir\'e and supermoir\'e on the participating dipolar excitons and, subsequently, the quadrupolar exciton has not been accounted for thus far, and is the aim of this work. 

\textbf{Fig. \ref{fig:1}a} illustrates the rich supermoir\'e landscape in a trilayer system, hosting different excitonic complexes governed by the local atomic registries. The heterobilayer regions (denoted as regions C, D) host the IX, whereas, in the trilayer region, the IXs in the two moir\'e potentials can couple differently depending on the local stacking. For example, at coordinates where the sites hosting $IX$ in the two moir\'e superlattices overlap (region A), the IXs can couple strongly and hybridize to form the $QX$. In contrast, at coordinates where these  moir\'e sites are not overlapping (region B), vertically staggered dipolar interlayer excitons will form. Interestingly, we expect that due to the formation of supermoir\'e structure, we shall always find aligned local registry around region A, even if the relative twist angles vary slightly (see \textbf{Supplementary section S1}). This weak dependence on the precision of the twist angle makes the observation of $QX$ more feasible in an experimental setup, which is in agreement with recent observation of $QX$ being robust against samples \cite{yuObservationQuadrupolarDipolar2023a}.
\par
We investigate a dual-gated \s/\se/\s~heterotrilayer (schematically shown in \textbf{Fig. \ref{fig:1}b}, optical image shown in \textbf{Supplementary section S2}) with all three layers edge-aligned to each other (see \textbf{Methods} for fabrication details). \colortext{We estimate the twist angle between top (bottom) WS$_2$ and WSe$_2$ to be 1.9$^{\circ}$ (0.7$^{\circ}$) by measuring the angle between sharp edges of individual flakes (see \textbf{Supplementary section S5}).} \textbf{Fig. \ref{fig:2}a} shows the representative photoluminescence (PL) of bilayer and trilayer regions at low-temperature ($\sim$ 4K) and zero field with a 633 nm excitation wavelength \colortext{(result from additional sample is shown in \textbf{Supplementary section S4}). This choice of excitation wavelength suppresses the initial electronic population in the top and bottom \s~layers as the excitation photon energy is lower than the \s~intra-layer free exciton. Thus, initially the charge carriers are only created in the \se~layer, followed by transfer of electrons to \s~layers due to type-II band alignment.}

The optically pumped excitonic density is kept well below the typical moir\'e density in order to avoid many body effects ( \textbf{Supplementary section {S3})}. The trilayer PL shows a clear redshift of $\sim$ 60 meV compared to the bilayer. We also observe a completely quenched \se~intra-layer exciton emission in the trilayer (\textbf{Supplementary section S4}), evidencing a high-quality interface with efficient charge transfer.

\colortext{Note that, doping-dependent dispersion of photoluminescence peaks of the trilayer region shows a filling behaviour at small gate voltage of $\pm$0.1V, where bilayer moir\'e filling is unlikely due to larger density. The estimated supermoir\'e density from the filling gate voltage is $4.98 \times 10^{11}$ cm$^{-2}$. On the other hand, the supermoir\'e density calculated from the independent twist angle measurement of the flake edges is $\approx 5 \times 10^{11}$ cm$^{-2}$ (\textbf{Supplementary section S5}). The close agreement of these two extracted density values from independent measurements supports the existence of supermoir\'e in our sample.}

Type-II band alignment between \s~and \se, as shown in \textbf{Fig \ref{fig:2}a (inset)}, ensures the formation of interlayer dipolar excitons with opposite polarity, IX$_u$ and IX$_d$ in the top and bottom heterobilayers, respectively. IX$_u$ and IX$_d$ will be trapped in the top and bottom moir\'e potential, respectively. The coordinates of the trapping site vary between the two moir\'e potentials due to unequal twist angles between constituting layers, leading to top and bottom moir\'e with different lattice constants and orientations. Importantly, IX$_u$ and IX$_d$ will only be vertically aligned when these sites overlap each other at points defined by the supermoir\'e (region A in \textbf{Fig. \ref{fig:1}a}). These specific sites, with strong coulomb coupling between IX$_u$ and IX$_d$, host the $QX$.
\par
In region A, due to the degenerate electronic state in the top and bottom \s~layers, electrons in the two layers can tunnel couple through the \se~barrier layer, leading to hybridization of the degenerate electronic states and formation of new symmetric and anti-symmetric electronic states. While the symmetric and anti-symmetric states are bright and dark\cite{slobodkinQuantumPhaseTransitions2020a}, respectively, at zero electric field, we show later in the text that the brightness of anti-symmetric states has high tunability with the applied electric field. This layer-hybridized electronic state can form a bound state with a hole confined in the \se~layer, thus leading to the formation of a $QX$ with a zero out-of-plane dipole moment. At zero electric field, the excitons funnel to the supermoir\'e points due to the coupling energy advantage (quantified by the emission energy difference between the bilayer and trilayer in \textbf{Fig. \ref{fig:2}a}), hence emission from supermoir\'e points dominates the spectra.
\par
We apply voltage bias at the two gates (\textbf{Fig. \ref{fig:1}b}) to create a vertical electric field ($F$) in the device while avoiding doping the sample (see \textbf{Methods}). \textbf{Figs. \ref{fig:2}b,c} illustrate the electric field-dependent PL measured from the bilayer and trilayer portions, respectively. The bilayer shows a linear Stark shift, as expected from a dipolar exciton. We extract a dipole moment of $\sim$ 0.5 e nm, consistent with previous reports on interlayer dipolar excitons \cite{montblanchConfinementLonglivedInterlayer2021, jaureguiElectricalControlInterlayer2019}. The trilayer shows a non-linear, hyperbolic Stark shift, directly evidencing the quadrupolar nature of emitting exciton \cite{yuObservationQuadrupolarDipolar2023a, xieBrightDarkQuadrupolar2023, liQuadrupolarDipolarExcitonic2023a}. This non-linear Stark shift originates due to the hybridization of the anti-parallel $IX_u$ and $IX_d$, and results in a hyperbolic field-dependent energy dispersion given by\cite{yuObservationQuadrupolarDipolar2023a, liQuadrupolarDipolarExcitonic2023a}
\begin{equation}
    E_{\pm}(F) = \mp \sqrt{(e.d.F)^2 + 
    \delta^2}
\label{eqn:energy}
\end{equation}
where $E_{+(-)}$ is the energy of a $QX$ with a symmetric (anti-symmetric) electronic state, e.d is the dipole moment of dipolar excitons being hybridized, and $\delta$ is the hybridization strength. We obtain a $\delta \sim$ 11.5 meV by fitting the electric field dependent energy at peak intensity with Eqn. \ref{eqn:energy} (\textbf{Supplementary section S6}). Note that the effect of hybridization will vanish at high field values ($F\gg \delta/ed$). This can be intuitively thought of as all electrons of the system being pushed to the top or bottom \s~layers depending on the sign of $F$ and the trilayer system effectively behaving like a bilayer system. Further, in the higher energy range (shown by the dashed box in \textbf{Fig. \ref{fig:2}c} with a zoomed-in view in log scale in \textbf{Fig. \ref{fig:2}d} for better visualisation), we observe a field induced blue shift of the emission energy, with the emission intensity increasing with the electric field - indicative of emission from the anti-symmetric $E_{-}$ branch (see \textbf{Supplementary section S7} for results at higher incident power showing conspicuous blue shift of $E_{-}$ branch with $F$). In addition, a careful observation of \textbf{Fig. \ref{fig:2}c} reveals a fine structure in the lower energy (symmetric) branch at nonzero $F$. The origin of these rich features in the light of moir\'e confinement and subsequent vertical moir\'e - moir\'e interaction is discussed below.

As illustrated in \textbf{Fig. \ref{fig:2}a}, the emission spectrum of the heterobilayer shows a three-peak structure, which aligns with recent reports of \s/\se~ moir\'e superlattices\cite{chatterjeeHarmonicAnharmonicTuning2023, tranEvidenceMoireExcitons2019}. The observed excitonic resonances are attributed to discrete energy levels due to confining harmonic moir\'e potential. As an important distinction, our heterotrilayer encompasses two distinct moir\'e potentials. At zero field, two degenerate sets of moir\'e confined excitonic states exist in the heterotrilayer. We denote these as $IX_{u(d)i}$ ($i=0,1,2$) for the $i^{th}$ excitonic state of the top (bottom) moir\'e, as schematically illustrated in the top panel of \textbf{Fig. \ref{fig:3}a}. The vertical alignment at the supermoir\'e sites results in strong Coulomb coupling between IX$_{ui}$ and IX$_{di}$. This causes the degenerate states to hybridize forming symmetric ($QX_{Si}$) and anti-symmetric ($QX_{Ai}$) quadrupolar states. 
\par
Note that the anti-symmetric states are dark at zero field due to vanishing electron-hole overlap and can be tuned to emit at a higher field. Thus, six quadrupolar states exist after the hybridization of top and bottom moir\'e, with half the quadrupolar states emitting and the rest are dark at small electric field values. The presence of only three bright excitonic states $QX_{S0}$, $QX_{S1}$ and $QX_{S2}$ at zero field in spite of two moir\'e systems being present in the system supports the inter-moir\'e hybridization hypothesis. We emphasize that the hybridization of the moir\'e states is independent of the detailed configuration and orientation of atomic layers, thanks to the supermoir\'e effect. Further, power-dependent measurement shows a monotonic increment in power law exponent from $QX_{S0}$ to $QX_{S2}$ (\textbf{Supplementary section S8}). This supports the moiré confined nature of QX, with a higher degree of confinement for lower energy states \cite{chatterjeeHarmonicAnharmonicTuning2023}.
\par
Due to the strong electric field tunability of the excitonic states in our sample, nonzero electric field can give rise to hybridization between $IX_{ui}$ and $IX_{dj}$ ($i\neq j$) states, leading to more complex quadrupolar states. Such interaction leads to anti-crossing between the symmetric (bright) and anti-symmetric (dark) states due to their field-dependent energy dispersion having opposite signs. Two such scenarios are schematically depicted in \textbf{Fig. \ref{fig:3}a} (bottom panel). Indeed, a closer examination of \textbf{Fig. \ref{fig:2}c} reveals a non-monotonic movement and anti-crossing of excitonic peaks at nonzero electric fields. Had there been no interaction between $IX_{ui}$ and $IX_{dj}$ ($i\neq j$), we would expect three emission peaks monotonically decreasing in energy with field.  
\par
To capture these effects, we construct the system's Hamiltonian as shown in Eq \ref{eqn:hamiltonian}, including interaction ($t_{ij}=t$) between $IX_{ui}$ and $IX_{dj(\neq i)}$. $\Delta_{m1} (\sim 20$ meV) and $\Delta_{m2} (\sim 40$ meV) correspond to the zero-field energy of the 2nd and 3rd electronic states with respect to the ground state (as extracted from the bilayer inter-peak separation in \textbf{Fig. \ref{fig:2}a}) due to energy quantization in harmonic moir\'e potential. The interaction terms $t_{i(=0,1,2)}$ account for the interaction between the antiparallel dipolar states IX$_{ui}$ and IX$_{di}$.
\begin{equation}
H = \begin{pmatrix}
-e.d.F & 0 & 0 & t_0 & t & t \\
0 & \Delta_{m1}-e.d.F & 0 & t & t_1 & t \\
0 & 0 & \Delta_{m2}-e.d.F & t & t & t_2 \\
t_0 & t & t & e.d.F & 0 & 0 \\
t & t_1 & t & 0 & \Delta_{m1}+e.d.F & 0 \\
t & t & t_2 & 0 & 0 & \Delta_{m2}+e.d.F 
\end{pmatrix}
\label{eqn:hamiltonian}
\end{equation}
\textbf{Fig. \ref{fig:3}b} shows the calculated field-dependent dispersion of the eigenvalues from the trilayer. The hybridization energy $t_i$, which is also the energy difference between the degeneracy broken states $QX_{Si}$ and $QX_{Ai}$, is estimated to be around $10 \sim 30$ meV \cite{slobodkinQuantumPhaseTransitions2020a}. On the other hand, the energy gap between the discrete states of the harmonic moir\'e well is around $20$ meV. Such an energy scale allows for different energy level ordering and crossing scenarios.
\par
\textbf{Fig. \ref{fig:3}b} exhibits several key features: (1) Gaps open at small field values due to anti-crossing between the highest-energy bright state and the lowest-energy dark state. This also leads to an unequal energy gaps between successive bright exciton resonances around zero-field. (2) We also observe that the third-lowest energy state (shaded in grey) now shows a non-monotonic field-dependent energy dispersion at low field values. \textbf{Fig. \ref{fig:3}c} shows the fitted field-dependent energies of the three observed peaks at small electric field values in the trilayer heterostructure from a finer electric-field dependent PL measurement (stacked spectra shown in \textbf{Supplementary section S9}). The color-coded guides in \textbf{Fig. \ref{fig:3}b-c} highlight that the experimental features are qualitatively well reproduced by our moir\'e-moir\'e interaction model. 
\par
Interestingly, we extract unequal slopes (effective dipole moment) of the three resonances in \textbf{Fig. \ref{fig:3}c}, as, $ed_{QX_{S0}}(= 0.59 \pm 0.01 \text{ e.nm}) > ed_{QX_{S1}} (= 0.42 \pm 0.02 \text{ e.nm}) > ed_{QX_{S2}} (= 0.23 \pm 0.04 \text{ e.nm})$. This observation of higher energy resonances exhibiting lower slope value (and hence stronger quadrupolar nature) is in agreement with the expectation of higher energy states having stronger tunnel coupling ($t_0<t_1<t_2$) (see \textbf{Supplementary section S10}). This is due to the difference in the net tunnel barrier experienced by the states with different indices, as illustrated in \textbf{Fig. \ref{fig:3}a}. The highest energy state has the lowest tunnel barrier. A lower tunnel barrier leads to a higher coupling constant and, thus, stronger hybridization.

We now focus on the higher energy antisymmetric $QX$ branch (\textbf{Fig. \ref{fig:2}c-d}). This branch shows a clear blue shift with the applied field (also see \textbf{Supplementary section S7} for results at higher power). This is contrary to the red shifting symmetric $QX$ resonances at the lower energy branch, and is indicative of the antisymmetric $QX$ branch. The intensity of these blue shifting peaks also increases with the magnitude of the applied electric field.
\par
To understand the field dependent brightness of $QX$, we calculate the electron-hole overlap for the symmetric and antisymmetric $QX$ (see \textbf{Supplementary section S11}). Following the idea of representing $QX$ as a linear combination of $IX$ states \cite{slobodkinQuantumPhaseTransitions2020a}, we find the electronic wavefunction in the heterotrilayer as a combination of two finite quantum wells with a type-II band structure inside each. \textbf{Fig. \ref{fig:4}a} shows the calculated electron-hole overlap values for the two branches (with the corresponding ratio being indicated in the inset). We find that the electron-hole overlap in antisymmetric $QX$ has a high degree of field tunability, dramatically enhancing from zero with increasing field strength. On the contrary, brightness of the symmetric state is found to be field tunable, but to a much smaller degree, decreasing with increasing field strength. 
\textbf{Fig. \ref{fig:4}b} shows the illustration of the mechanism for brightness tunability in quadrupolar states. The size of the electronic wavefunctions in the illustration is scaled according to the respective amplitude in the quadrupolar state, with a larger shape symbolizing greater contribution magnitude compared to the other dipolar state and vice versa. Starting off at zero field, the anti-parallel $IX$ states contribute equally to $QX$. However, due to opposite phases of $IX$ states in the $QX_{A}$ state, the net electron-hole overlap is found to be zero\cite{slobodkinQuantumPhaseTransitions2020a}. As the applied electric field increases, the electron in the dipolar state, parallel to the electric field, with a higher contribution to $QX_{S}$, localizes farther from the hole wavefunction due to band bending, resulting in reduced overlap. In sharp contrast, the high energy dipolar state, anti-parallel to the electric field, has a higher contribution in $QX_{A}$. The electron wave function of the high energy IX state localizes closer to the hole wavefunction, resulting in higher overlap and, consequently, brightening of the $QX_{A}$ state. 
\par
As we argued earlier, the top and bottom moir\'e interact strongly only at certain locations in the lattice (region A in \textbf{Fig. \ref{fig:1}a}), defined by the supermoir\'e. At low electric field, excitons funnel to these points due to energy advantage through forming a $QX$. However, at higher electric field, the energy advantage by forming a $QX$ is negligible, suppressing exciton funnelling to the supermoir\'e points, as schematically illustrated in \textbf{Fig \ref{fig:4}c}. Hence the dipoles from region B (defined in \textbf{Fig. \ref{fig:1}a}) also start contributing appreciably to the total emission at high electric field, along with region A. The vertical interaction between the top and bottom moir\'e in region B is negligible due to spatially staggered nature. Thus, both the top and bottom moir\'e pockets emit at a higher electric field, with one of them at a higher and the other at a lower emission energy due to the anti-parallel and parallel electric field, respectively. The emission intensity of the higher energy states increases with the field due to enhanced electron-hole overlap, as discussed above. As the funneling quenches and all moir\'e points start emitting at high field (in contrast to only the supermoir\'e points emitting at lower field), the effective emission area on the sample (and hence inhomogeneity) also increases. The emergence of multiple peaks at higher electric field is a signature of this effect (see \textbf{Supplementary section S12}). Exciton funneling effect to supermoir\'e sites is further evidenced by the dependence of power induced blueshift on the applied electric field in  \textbf{Supplementary section S13}. The brightening of these states at high field values, combined with the spectral position and field-induced blue-shift, establishes the attribution of these peaks as $QX_A$ states. {Finally, we observe a field and energy dependent polarity switching of the residual polarization (likely arising due to slight asymmetry between the top and bottom moir\'e) - suggesting the lower and higher energy peaks originate from the $QX_S$ and $QX_A$ branches of the quadrupolar exciton (see \textbf{Supplementary section S14}).}
\section{Methods}
\textbf{\label{subsec:fab}Device Fabrication.} The heterotrilayer \s/\se/\s~is top and bottom capped with hBN and with few-layer graphene as a top gate. The device is fabricated using a sequential dry-transfer method (with micromanipulators). The individual monolayers are aligned along crystallographic axes using straight edges. The heterostructure is prepared on top of a pre-patterned Ti/Au (10/50 nm) electrode, which is used as a back gate. The \s~and \se~layers are contacted to a grounded graphite electrode for carrier injection. To enhance the adhesion between layers, the devices are annealed inside a vacuum chamber ($10^{-6}$ mbar) at 150 $^\circ$C for 3 h after completion of transfer process.

\textbf{PL Measurement.} The PL measurements are carried out in Montana cryostat at 4 K. A $\times$50 objective lens with NA = 0.5 is used for excitation and collection in a back-reflection geometry. The PL spectra are collected using a spectrometer with 1800 lines per mm grating. The top and bottom gate voltages are applied using a Keithley 2636B source meter. A 633 nm CW laser (spot size $\sim$ 1.5 $\mu$m) is used to excite the sample with $6$ $\mu$W excitation power unless mentioned otherwise.

\textbf{Electric Field in the Heterostructure.} The electric field (F) inside the heterostructure is calculated by assuming a parallel-plate capacitor mode of the heterostructure. We get,
\begin{equation*}
    F = \frac{\epsilon_{hBN}}{\epsilon_{TMD}}\left(\frac{V_T}{d_T} - \frac{V_B}{d_B}\right)
\end{equation*}
where $V_{T}$ and $V_{B}$ are the top and bottom gate voltages, respectively, $d_T$ ($\simeq$ 6.5 nm) and $d_B$ ($\simeq$ 13 nm) are the thickness of the top and bottom hBN, respectively. $\epsilon_{hBN}$ (= 3.9) and $\epsilon_{TMD}$ (= 7.2) are the relative permittivities of hBN and TMDs, respectively.
\section*{Supporting Information}
The Supporting Information is available free of charge at -

Robustness of supermoir\'e, optical image of device, estimation of excitonic density, raw trilayer photoluminescence spectra and results from additional sample, twist angle and supermoir\'e effect, estimating hybridization energy in trilayer, field dependence of quadrupolar exciton at higher power, power law of different resonances, finer resolution trilayer electric field dependence, effect of coupling constant on the energy-field slope (dipole moment) at high field, conduction band hybridization and field dependent overlap of electron and hole wavefunction, anti-symmetric branch at high electric field showing multiple peaks, power dependence at different electric fields, and polarization resolved trilayer emission.

\section*{Acknowledgements}
A.K. acknowledges support from the INSPIRE Scholarship for Higher Education (SHE) from the Department of Science and Technology, Government of India. K.M. acknowledges the support from National Quantum Mission, an initiative of the Department of Science and Technology (DST), Government of India, a grant under SERB TETRA, grants from Indian Space Research Organization (ISRO), a grant from British Telecom India Research Centre (BTIRC), a grant from I-HUB QTF, IISER Pune, and a seed funding under Quantum Research Park (QuRP) from Karnataka Innovation and Technology Society (KITS), K-Tech, Government of Karnataka. K.W. and T.T. acknowledge support from the JSPS KAKENHI (Grant Numbers 21H05233 and 23H02052) and World Premier International Research Center Initiative (WPI), MEXT, Japan. 

\section*{Competing Interests}
The authors declare no competing financial or non-financial interests.
\section*{Data Availability}
Data available from the corresponding author on reasonable request.
\bibliography{n_references}
\clearpage

\begin{figure}
\centering
\includegraphics[width=\textwidth]{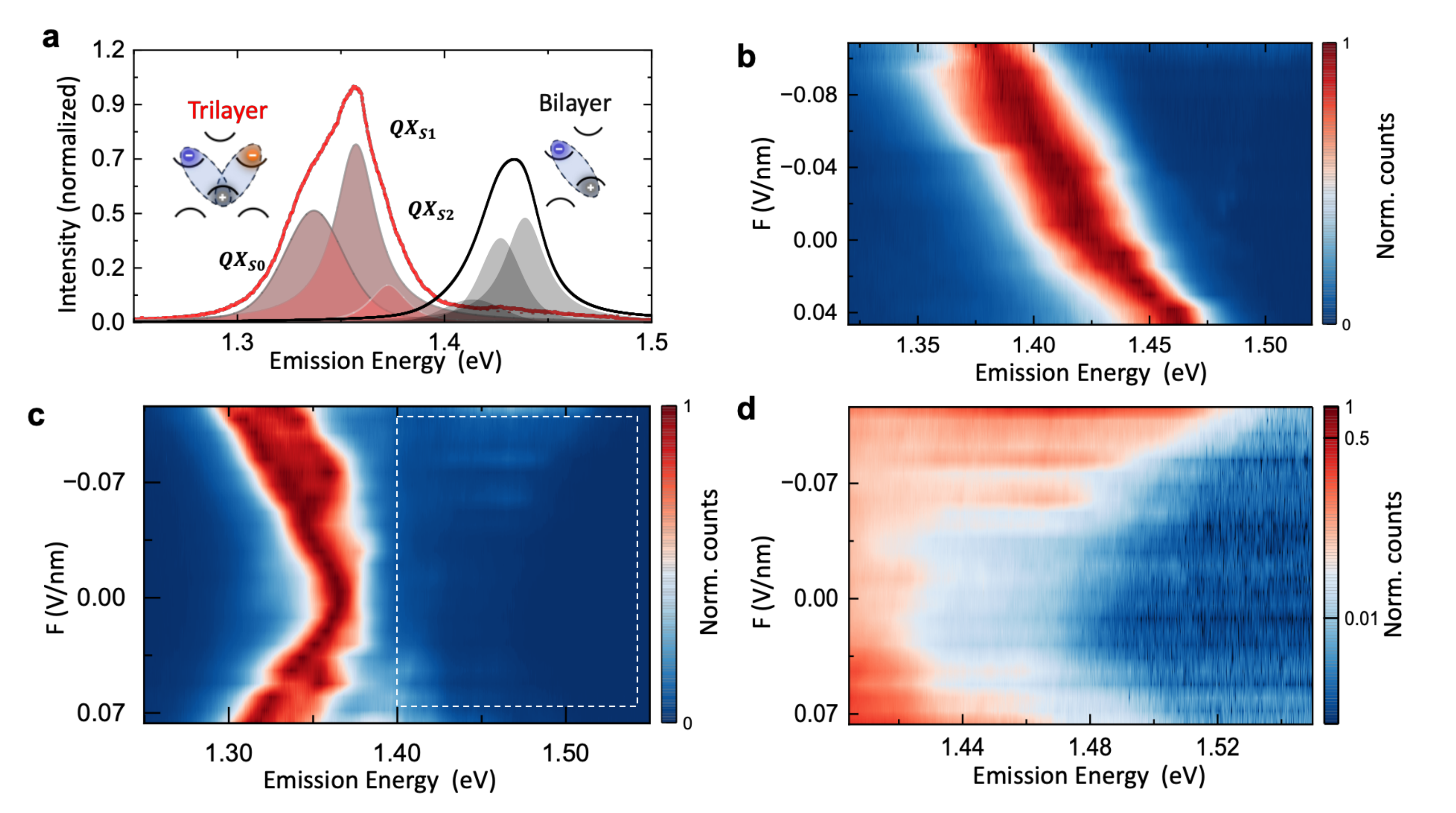}
\caption{\textbf{Emission features of dipolar and quadrupolar exciton:}\textbf{(a)} Photoluminescence (PL) Spectra from \s/\se/\s~ trilayer and \s/\se~ bilayer regions of the heterostructure with their corresponding decomposed peaks. The insets illustrate the simplified band structure of the corresponding regions. \textbf{(b,c)} Colour plot of electric field dependent PL for \textbf{(b)} \s/\se~ bilayer and \textbf{(c)} \s/\se/\s~ trilayer. \textbf{(d)} Colour plot of the higher energy emission (corresponding to the white dashed box in \textbf{c}) from trilayer region in log scale. {All spectra in (b,c,d) are individually normalized to 1.}}
\label{fig:2}
\end{figure}

\begin{figure}
\centering
\includegraphics[width=\textwidth]{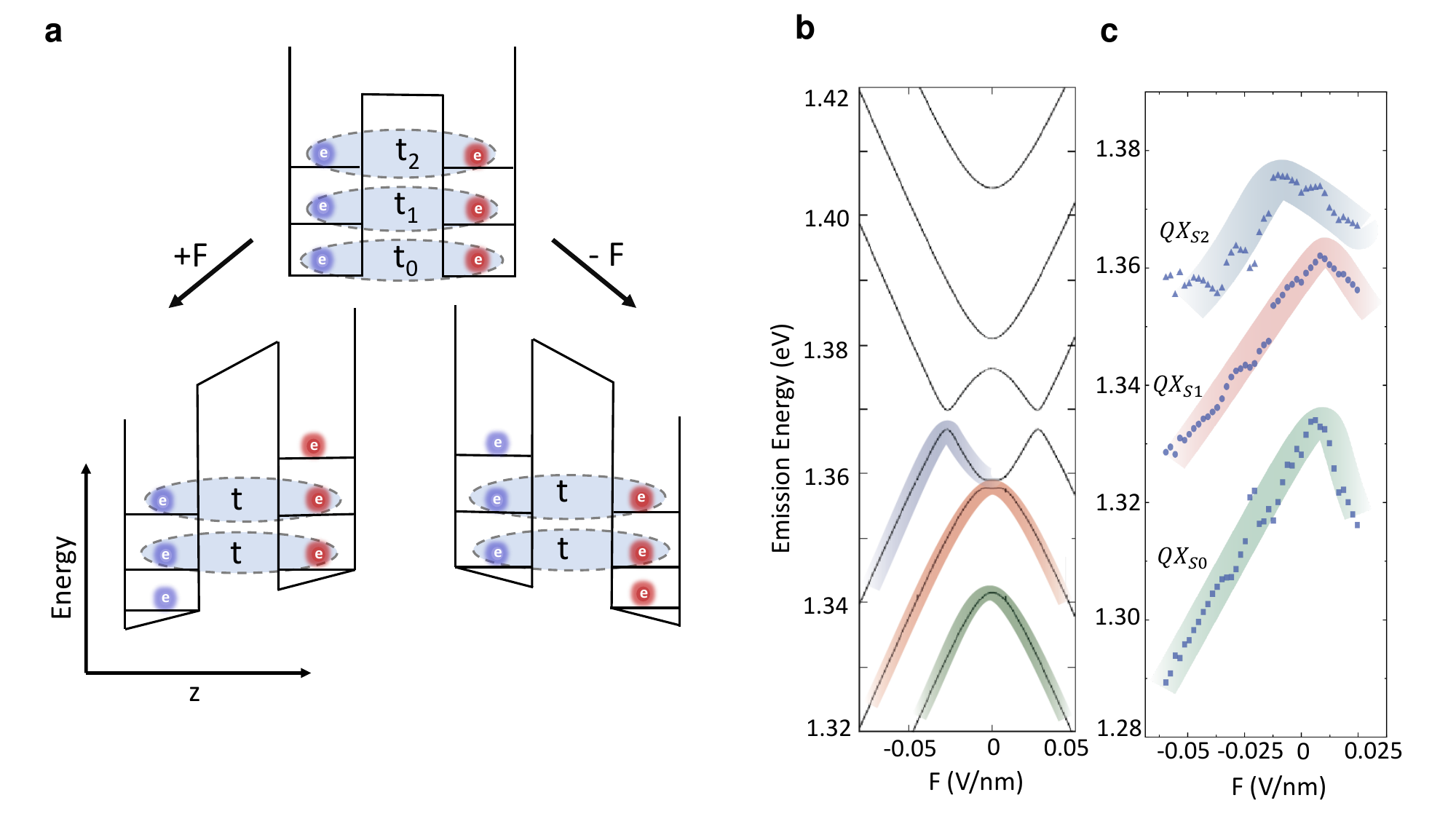}
\caption{\textbf{Vertical moir\'e-moir\'e interaction:} \textbf{(a)} Simplified potential well picture for the two vertically-coupled moir\'e wells. (top) The situation at zero applied electric field. The corresponding $t_{i=\{0,1,2\}}$ represents the coupling between the electronic states between the top and bottom \s layers with the same level in the moir\'e wells. (bottom) At nonzero electric field, $t$ represents coupling between states $QX_{i \neq j}$.  \textbf{(b)} Calculated and \textbf{(c)} measured electric field dependence for the energy of $QX$ states. The colored overlays show the correspondence between predicted and observed dispersions. The emission energy in \textbf{(b)} is offset to match the measured emission energy.}
\label{fig:3}
\end{figure}

\begin{figure}
\centering
\includegraphics[width=\textwidth]{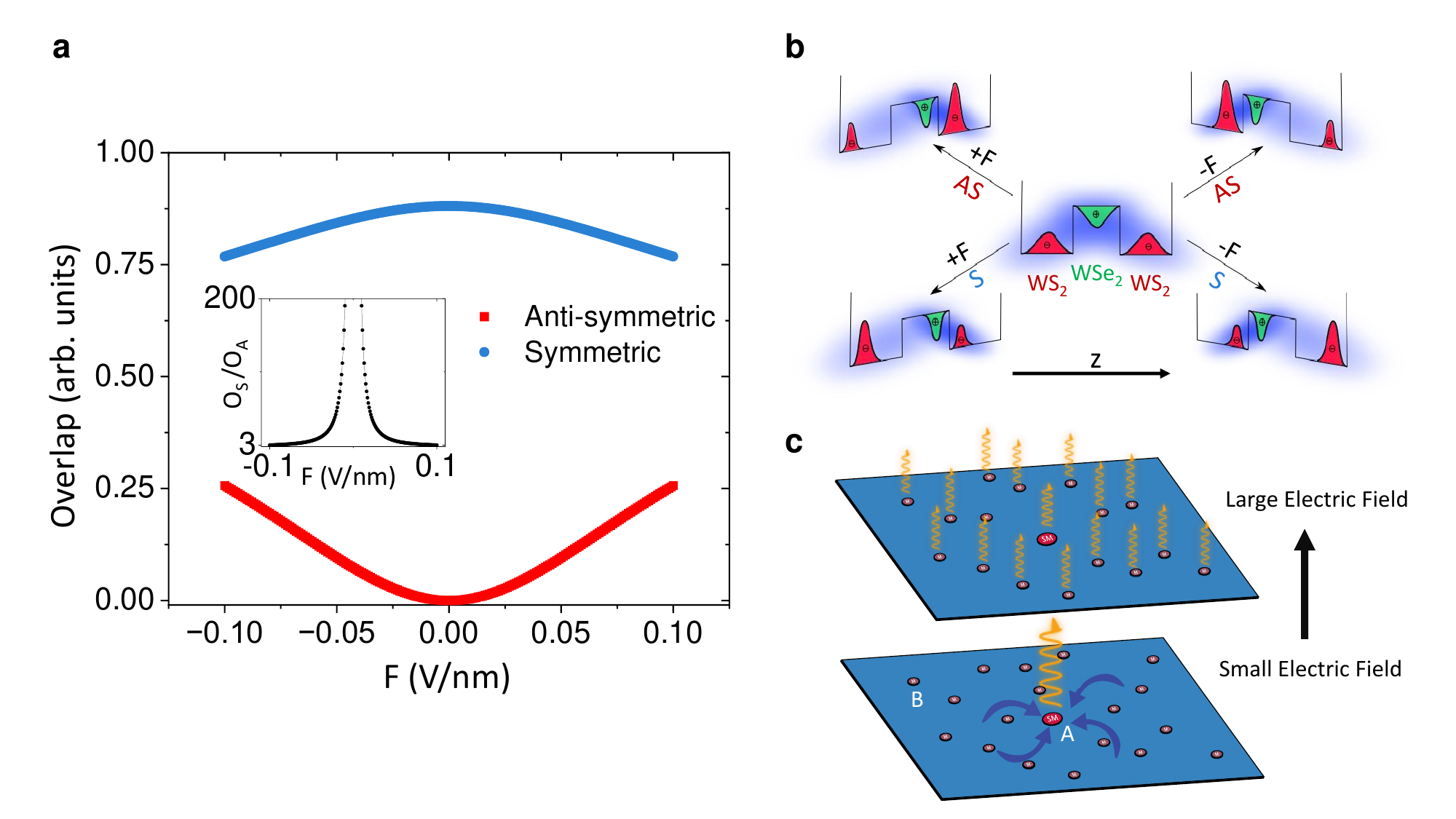}
\caption{\textbf{Quadrupolar exciton at high electric field: }\textbf{(a)} External electric field dependence of calculated overlap between the electron and hole wavefunctions in symmetric and anti-symmetric $QX$s. The inset shows the change in ratio of electron-hole wavefunction overlap for symmetric ($O_S$) and anti-symmetric ($O_{A}$) state with external electric field. \textbf{(b)} Illustration of the effect of external electric field on symmetric and antisymmetric $QX$. The comparative sizes of the electron wave packets illustrate the weight of the electronic wavefunction in the corresponding potential well. The color of the hue connecting the wave packets represents the overlap strength (darker meaning more overlap.) \textbf{(c)} At small electric field, funneling of the excitons into supermoir\'e point (large red circle) to form quadrupolar exciton due to energy advantage, followed by emission from the the supermoir\'e  point. At higher electric field, the energy advantage is suppressed, and the excitons emit from larger area (small orange circle).}
\label{fig:4}
\end{figure}
\newpage
\begin{figure}
\centering
\includegraphics[width=\textwidth]{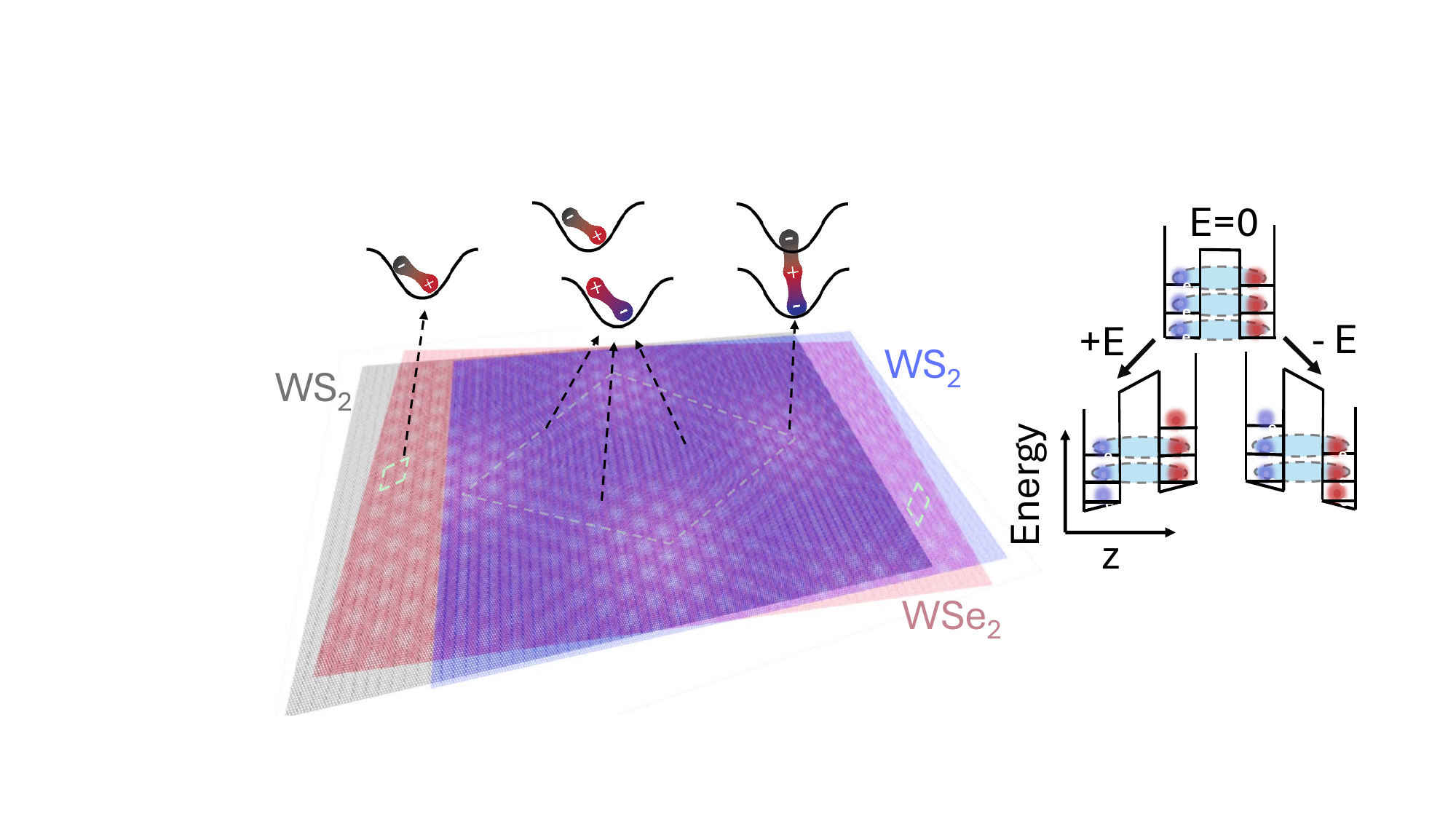}
\caption*{TOC graphic}
\end{figure}


\section*{Supplementary material}

# Supporting Information: Supermoiré-trapped quadrupolar exciton


Anish Kumar,† Suman Chatterjee,† Kenji Watanabe,‡ Takashi Taniguchi,¶ and Kausik Majumdar*,†

†*Department of Electrical Communication Engineering, Indian Institute of Science, Bangalore, India, 560012*

‡*Research Center for Electronic and Optical Materials, National Institute of Material Science, 1-1 Namiki, Tsukuba, 305-0044, Japan*

¶*Research Center for Materials Nanoarchitectonics, National Institute of Material Science, 1-1 Namiki, Tsukuba, 305-0044, Japan*

E-mail: kausikm@iisc.ac.in

# Supporting Information

## S1 Robustness of supermoiré

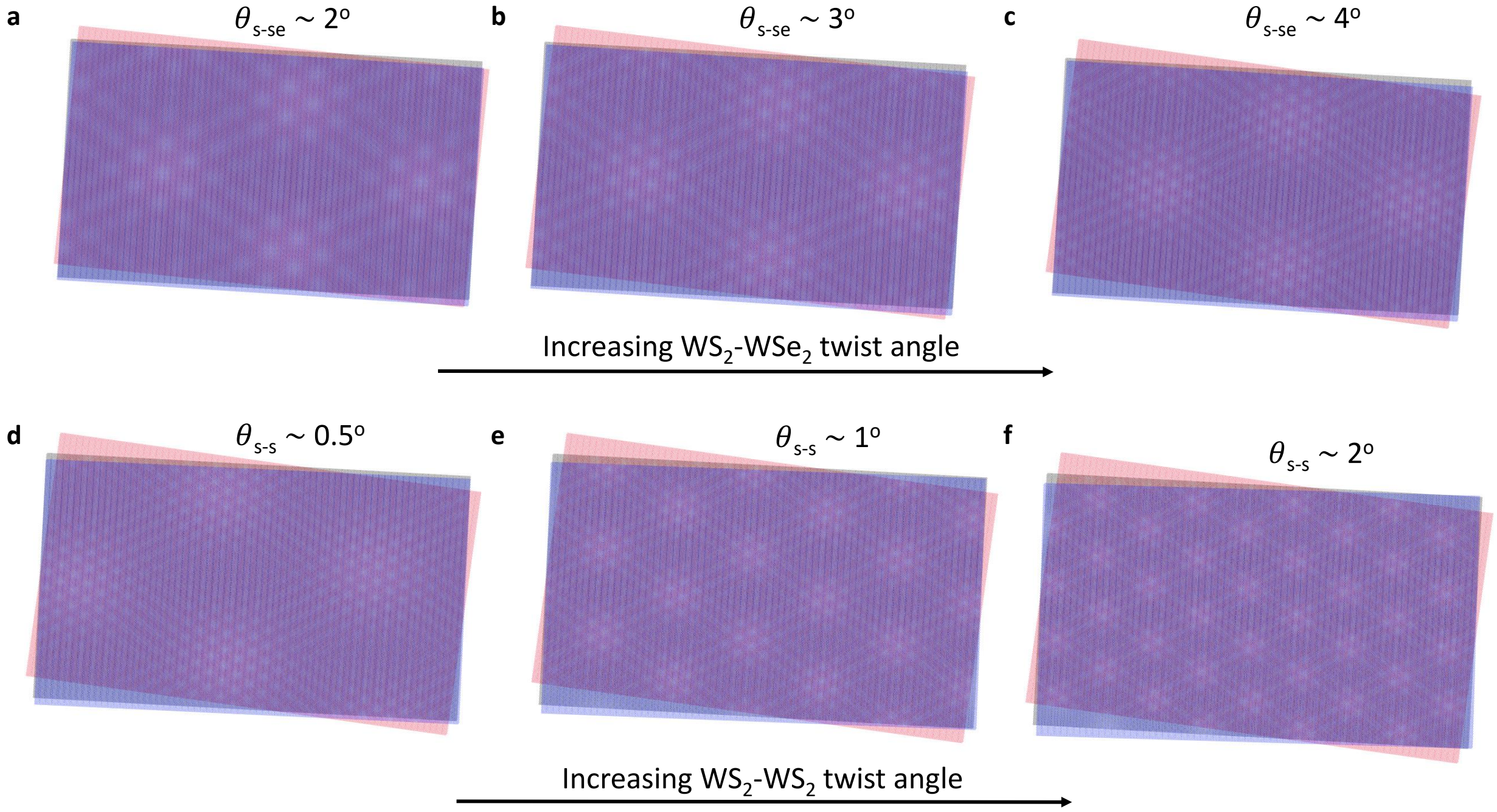


Figure SF1: **(a,b,c)** Different twist angles between the $WSe_2$ and $WS_2$ ($\theta_{s-se}$) layers while keeping the angle between $WS_2$ layers fixed. Changing $\theta_{s-se}$ changes the density of smaller moiré into the bigger supermoiré while length scale of supermoiré remains fixed. **(d,e,f)**. Different twist angles between the top and bottom $WS_2$ layers ($\theta_{s-s}$) layers without changing the angle of $WSe_2$. Since the twist between the two $WS_2$ layers defines the supermoiré, change in $\theta_{s-s}$ changes the length scale of supermoiré. The distance between different moiré sites is changing, which can result in different quadrupole-quadrupole interaction for devices with varying $\theta_{s-s}$ twist angle. The average area covered by supermoiré points is found to be similar across different angles. The consistency in formation of supermoiré across different twist angles reflects the robustness of quadrupole exciton formation against twist angle.

## S2 Optical image of device

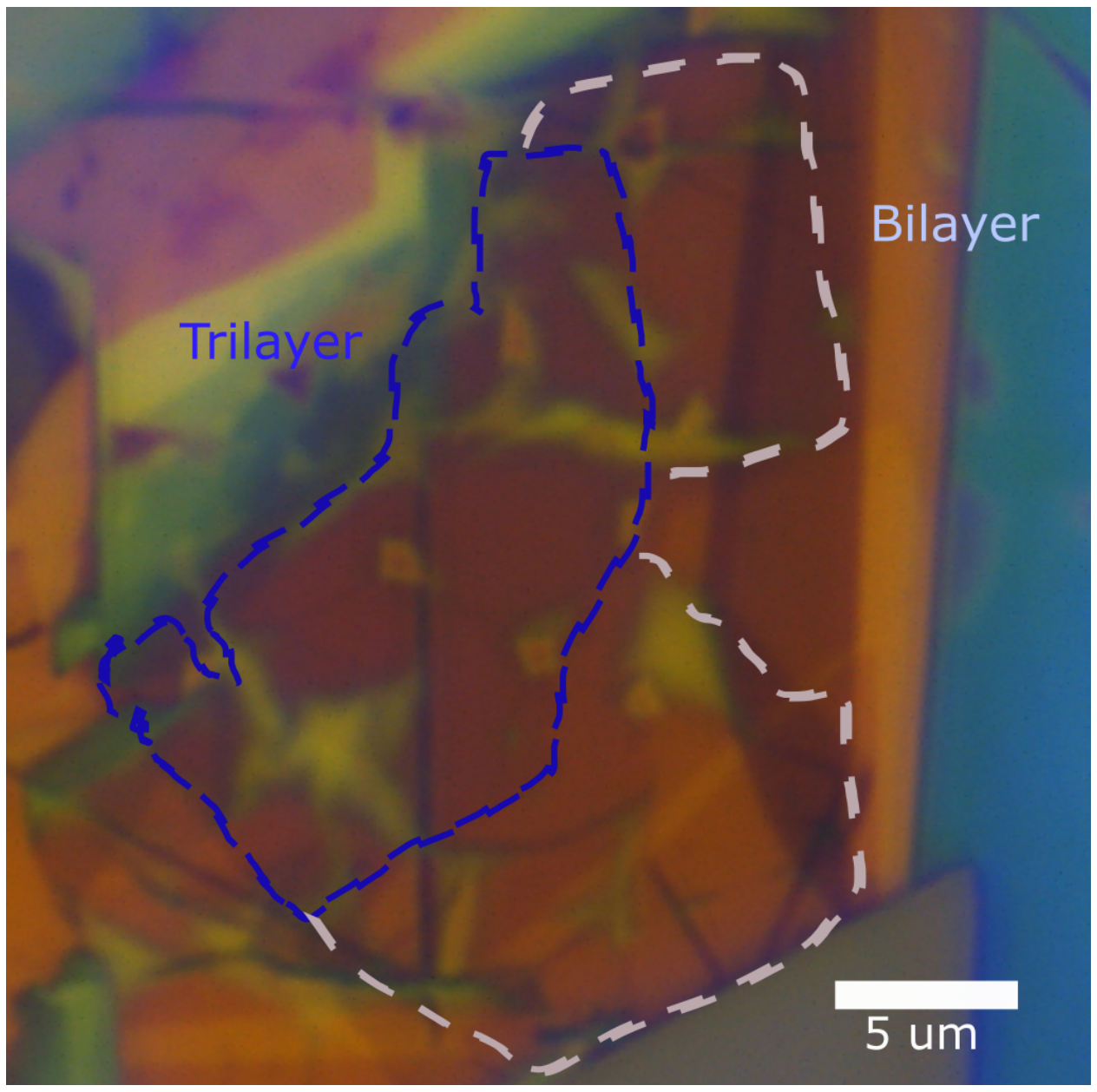


Figure SF2: Optical image of the dual gated and hBN encapsulated angle-aligned $WS_2/WSe_2/WS_2$ heterostructure. The dashed line encloses the heterostructure area.

## S3 Estimation of excitonic density

High excitonic density, exceeding the moiré pocket density can lead to involvement of many-body interactions, confounding the bare quadrupole physics. We note that the power mentioned in the main text is the output power of the laser. We measured the loss in excitation pass to be 49%. The on-sample excitation power is thus around 6 $\mu$W. While calculating the optically pumped excitonic density of interlayer excitons, absorbance and quantum efficiency of charge separation are the major source of uncertainty in the estimation. For this estimation, we assume a 0.5% absorbance of the trilayer[1] along with 100% quantum efficiency of excitation to quadrupole formation due to the ultra-fast charge transfer[2] (to obtain an upper limit of exciton density). We obtain:

$$N_g = \left\{ \frac{f \times P}{\pi r^2 \times \hbar\omega} \right\} = 5 \times 10^{18} \mathrm{cm}^{-2} \cdot \mathrm{s}^{-1}$$

and,

$$N_s = N_g \times \tau = 5 \times 10^{10} \mathrm{cm}^{-2}$$

where $N_s$, $N_g$ and $\tau$ are steady state excitonic density, rate of generated excitonic density and lifetime, respectively. $f$ is the absorbance of the heterostructure, $P$ is the optical power, $r$ is the beam radius and $\omega$ is the laser frequency.

Given the 1.5 $\mu$m beam diameter, 633 nm excitation and  10 ns measured lifetime of the quadrupole, we estimate the steady state excitonic density to be of the order $10^{11}$ cm$^{-2}$. Comparing this to the estimated moiré density of $3 \times 10^{12}$ cm$^{-2}$, the optically pumped exciton density is an order of magnitude smaller than the moiré density and thus exciton-exciton interaction can be neglected.

## S4 Raw trilayer photoluminescence spectra and results from additional sample

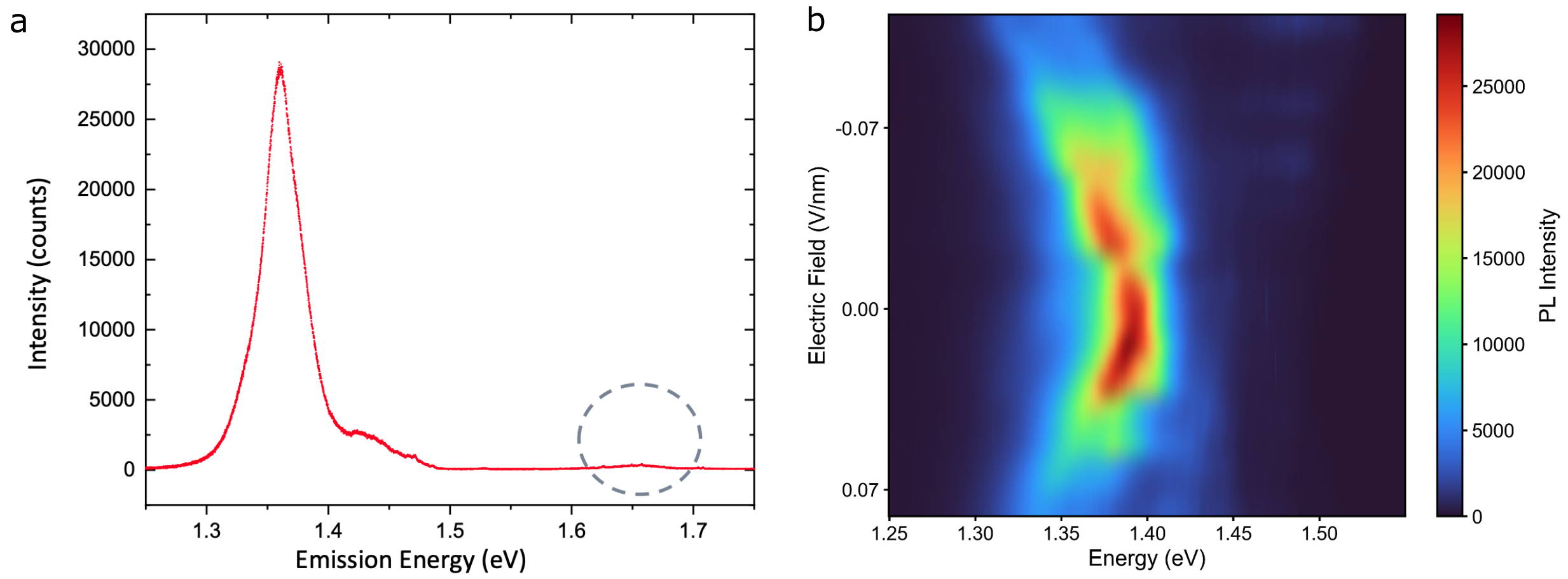


Figure SF3: Additional photoluminescence spectra of trilayer. **(a)** Suppressed emission from $WSe_2$ intralayer excitonic complexes (around 1.65-1.7 eV, highlighted with dashed circle) in trilayer suggests enhanced coupling between $WS_2$ and $WSe_2$ layers. The electron transfer efficiency from $WSe_2$ layer is higher in trilayer due to $WSe_2$ capping from both sides, leading to suppressed intralayer exciton emission from $WSe_2$ layer. **(b)** Electric field dependent raw spectra (without normalization) of trilayer. Oscillator strength of the PL decreases with magnitude of electric field due to reducing electron-hole overlap in symmetric quadrupolar exciton. The individually normalized field-dependence is shown in main text Fig. 2(c).

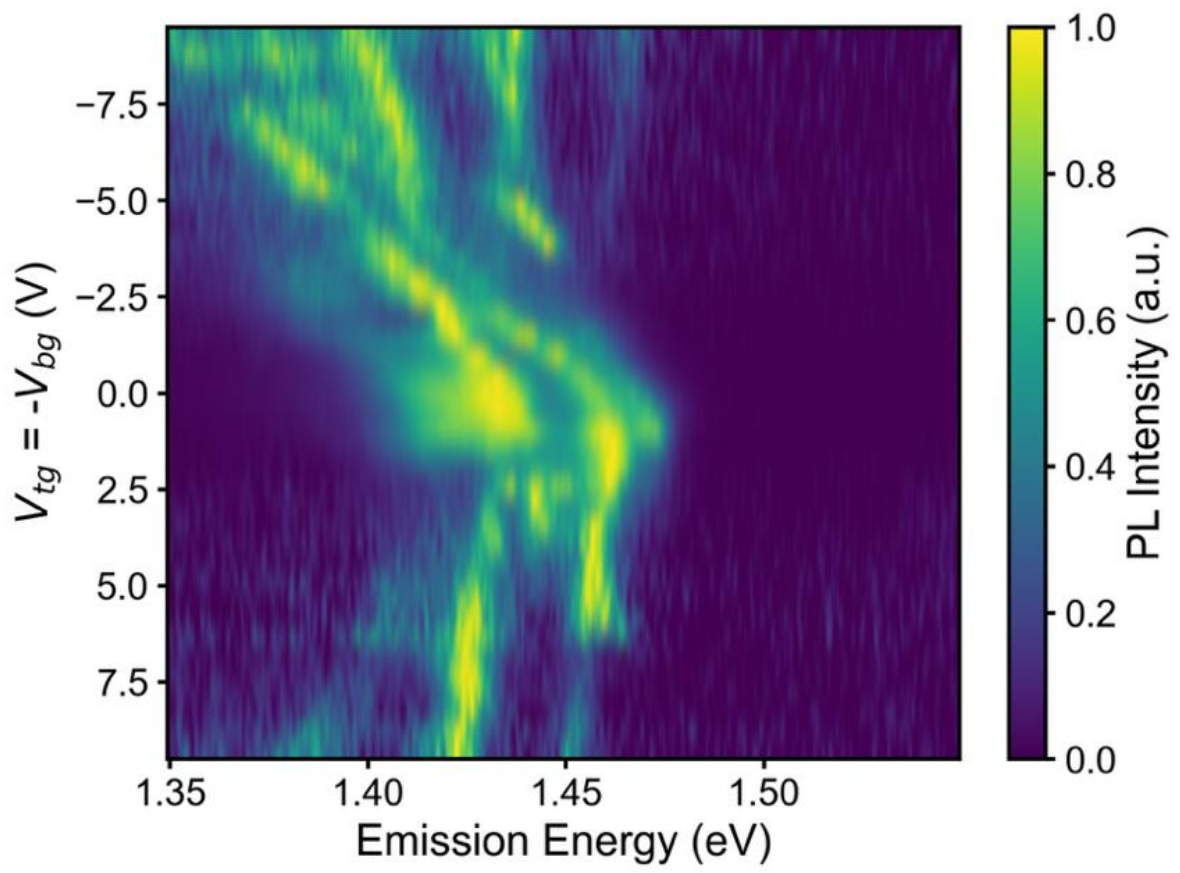


Figure SF4: Electric field dependent normalized spectra from another trilayer sample, showing the dispersion and interaction of different excitonic branches.

## S5 Twist angle and supermoiré effect

We determine the twist angle from the optical image of the sample in order to compute the supermoiré density. We calculate a twist angle of 1.9° (0.7°) for top $WS_2$ (bottom $WS_2$) and $WSe_2$ (see Fig. SF5). To obtain a supermoiré period from interference of these two moiré superlattices, we use the equation below:[3]

$$\lambda_{SM_1} = \frac{(1+D)\lambda_t}{\sqrt{2(1+D)(1-cos\phi)+D^2}}$$

$$D = \frac{|\lambda_b - \lambda_t|}{\lambda_t}, tan\phi_{t,b} = \frac{-sin\theta_{t,b}}{(1+\delta)-cos\theta_{t,b}}$$

Here, $\delta$ is the lattice mismatch between $WS_2$ and $WSe_2$, $\lambda_b$ ($\lambda_t$) is the moiré period between the bottom (top) and middle $WSe_2$ layer, and $\theta_t(\theta_b)$ is the twist angle between the bottom (top) and middle layer.We calculate a supermoiré period and density of 15.6 nm and $4.7\times10^{11}$ $cm^{-2}$ respectively, using this equation.

Also, we calculate the moiré period between top and bottom $WS_2$ by measuring the twist angle ($\theta = 1.2°$). Here, we use the equation $\lambda_{SM_2} = a/\theta$,[4,5] (a = 0.31 nm). Considering interference, this approximately gives us a supermoiré period of 15.1 nm with a similar density of $5.1 \times 10^{11}$ $cm^{-2}$ as before.

We measure the doping dependent PL on the trilayer which shows peaks in oscillator strength at different doping levels, as shown in the **Fig. SF6a**. Change in the oscillator strength at fractional fillings show the existence of moiré in the heterostructure. We repeated the measurement at two different spots (**Fig. SF6b**) and observed peaks in intensity around similar voltages, as shown in the figure below. This shows that the peaks are not due to local inhomogeneity. We assign each intensity peak on the hole side to half filling of the moiré bands. This gives us an estimated twist angle of 2° with a moiré superlattice constant of $\approx$ 5.8 nm and moiré density of $3 \times 10^{12}$ $cm^{-2}$.

Interestingly, we observe peaks very close to zero doping (±0.1V). No fractional filling

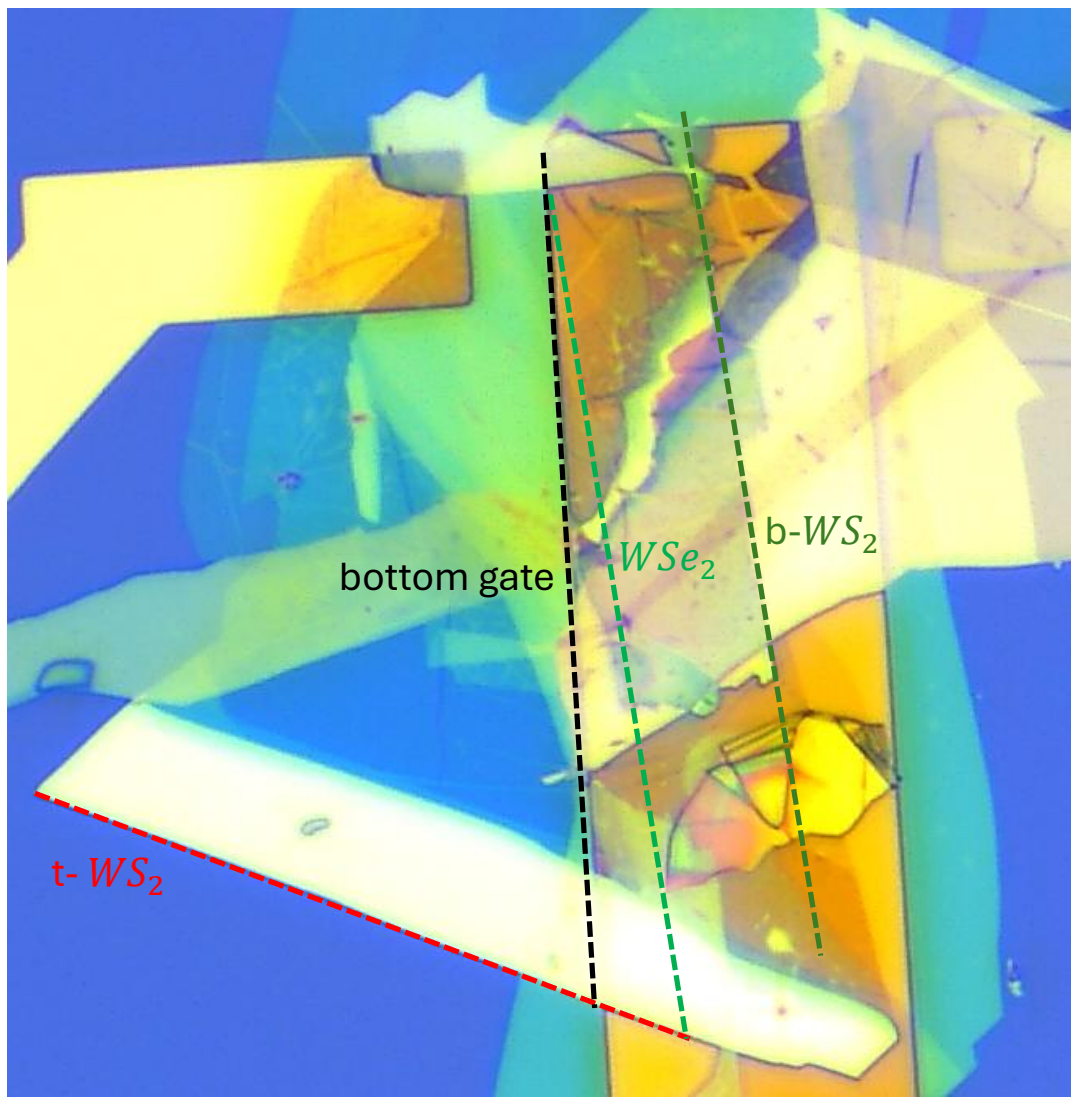


Figure SF5: Optical image of the sample showing flake edges (marked by colored dashed lines). All twist angles are measured with respect to the bottom gate edge (black dashed).

effects of this magnitude is expected for a moiré lattice. We attribute this feature to the filling of supermoiré lattice in the trilayer heterostructure. The features are observed at both the spots on the sample. We calculate the expected doping density with measured hBN thickness values (see Methods) of $d_T$ $(d_b) = 6.5$ (13) nm, which yields $4.98 \times 10^{11}$ cm$^{-2}$. This is in excellent agreement with the independently calculated supermoiré density from the measured twist angles (Fig. SF5). This agreement verifies the existence of the supermoiré lattice.

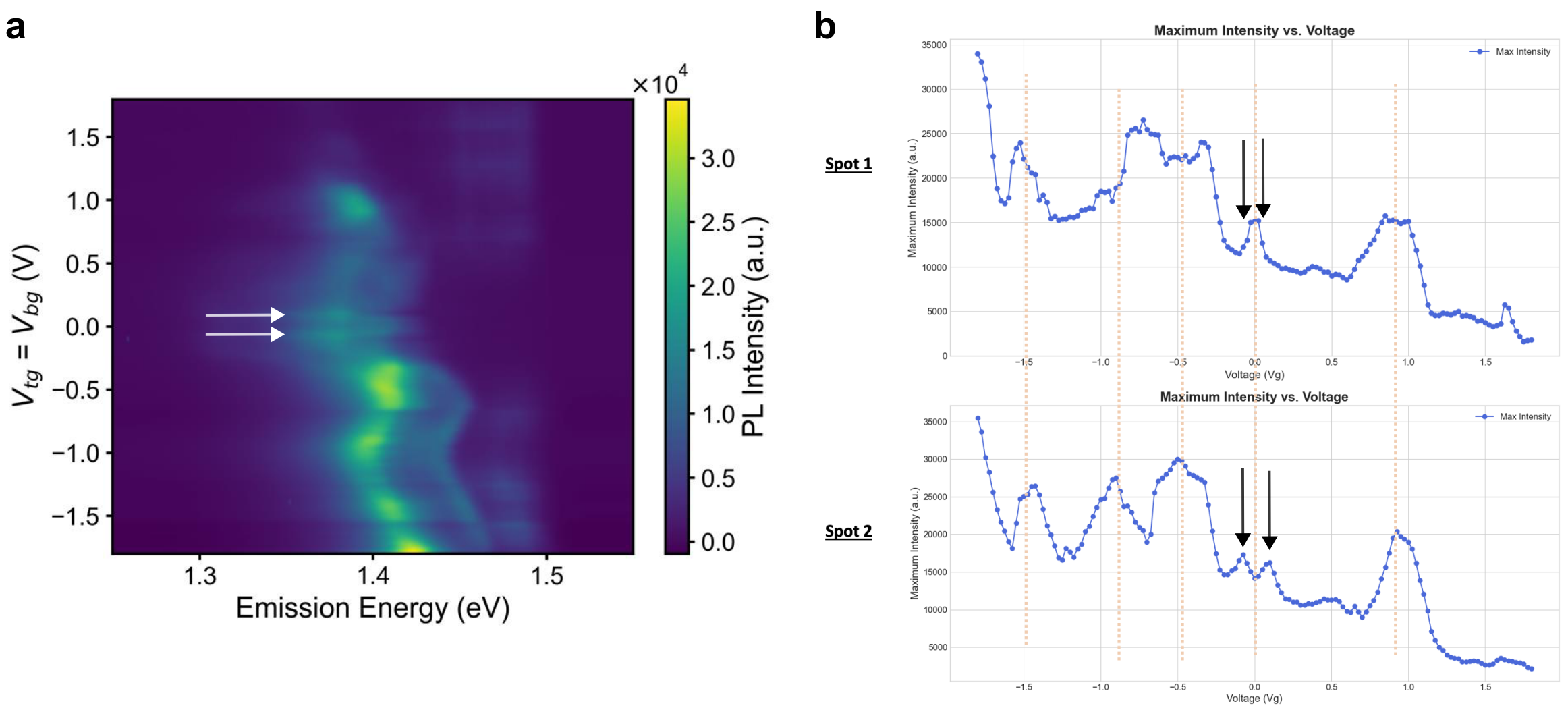


Figure SF6: **(a)** Doping dependent dispersion of PL peaks of the trilayer. (b) Doping dependent integrated PL of the trilayer at two different spots on the trilayer. The dashed orange lines are a guide to show correspondence between the two spots. Spot 2 corresponds to the data shown in (a). The arrows in (a,b) show PL enhancement on minute electron and hole doping.

## S6 Estimating hybridization energy in trilayer

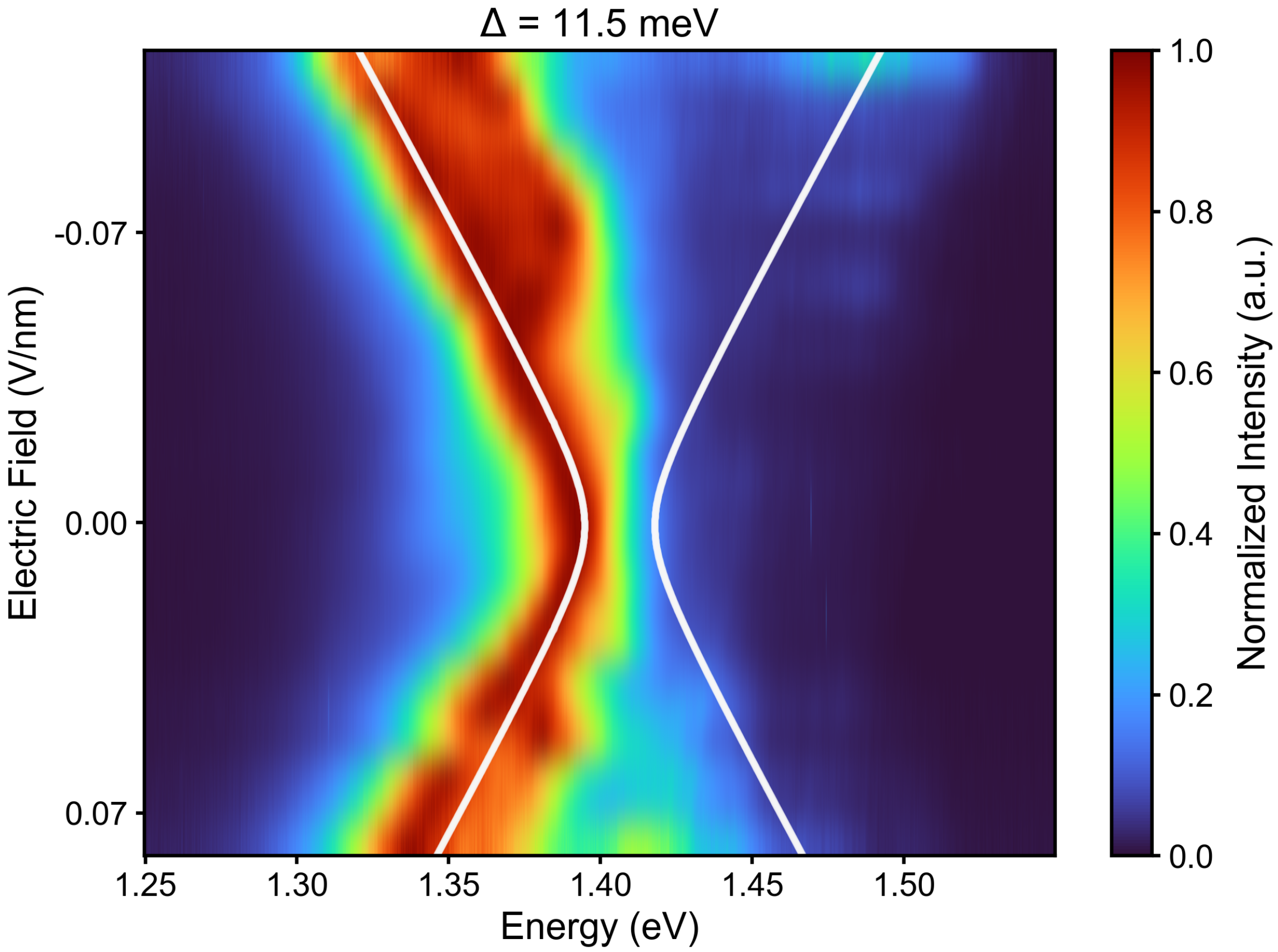


Figure SF7: Electric field dependent energy at peak intensity is fitted with Eqn. 1 in the main text to estimate the hybridization energy of the quadrupolar exciton. The solid white line on the left shows the fitting for the symmetric QX. The solid line on the right shows the expected dispersion of the anti-symmetric branch of the QX. This fitting matches the emission energies at higher energies under large electric field.

## S7 Field dependence of quadrupolar exciton at higher power

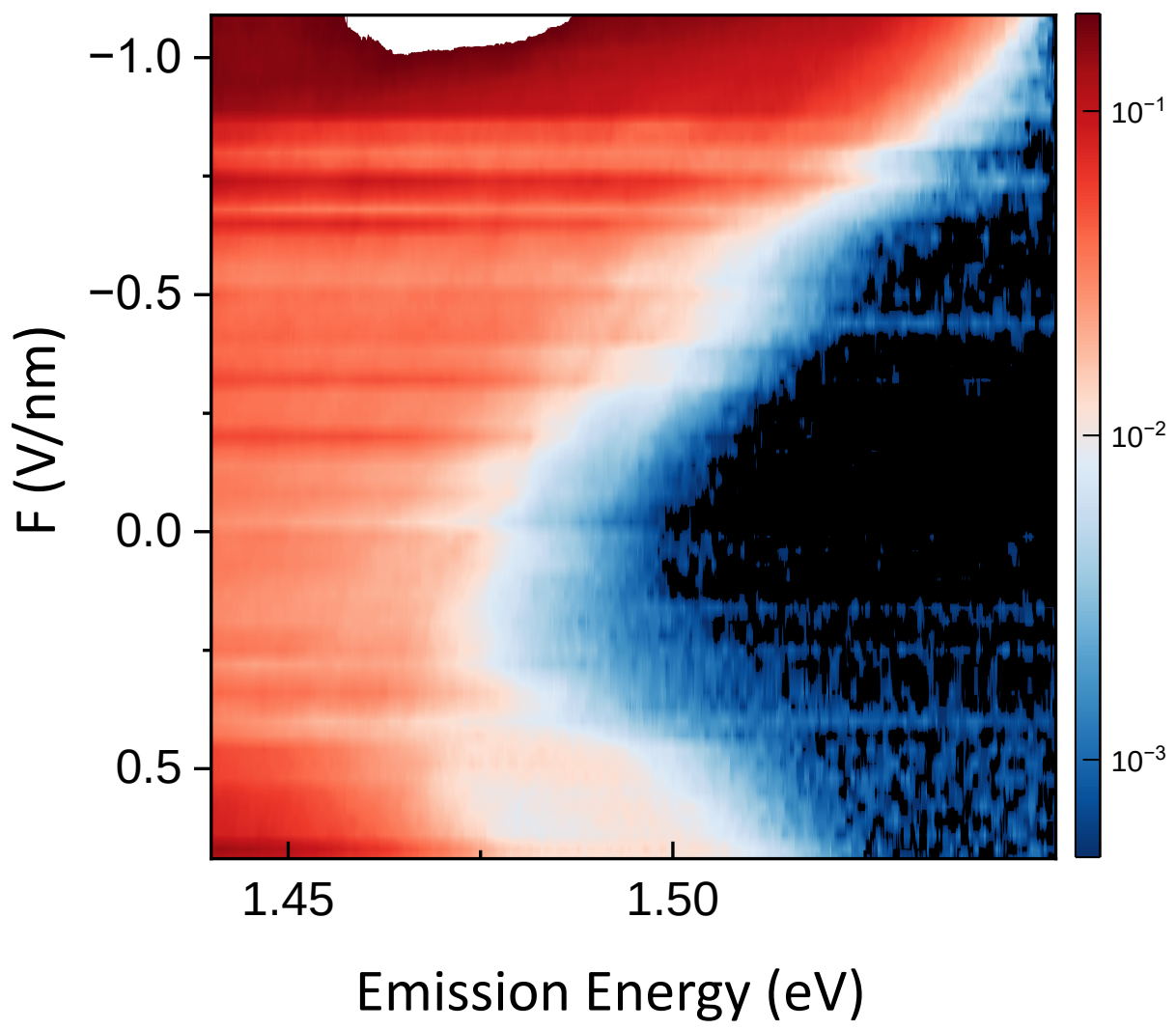


Figure SF8: Electric field dispersion of the high energy (anti-symmetric) branch of the quadrupolar exciton (plotted in logarithmic scale) at higher excitation power (320 $\mu$W), exhibiting prominent blue shift with an increase in the absolute value of the vertical electric field.

## S8 Power law of different resonances

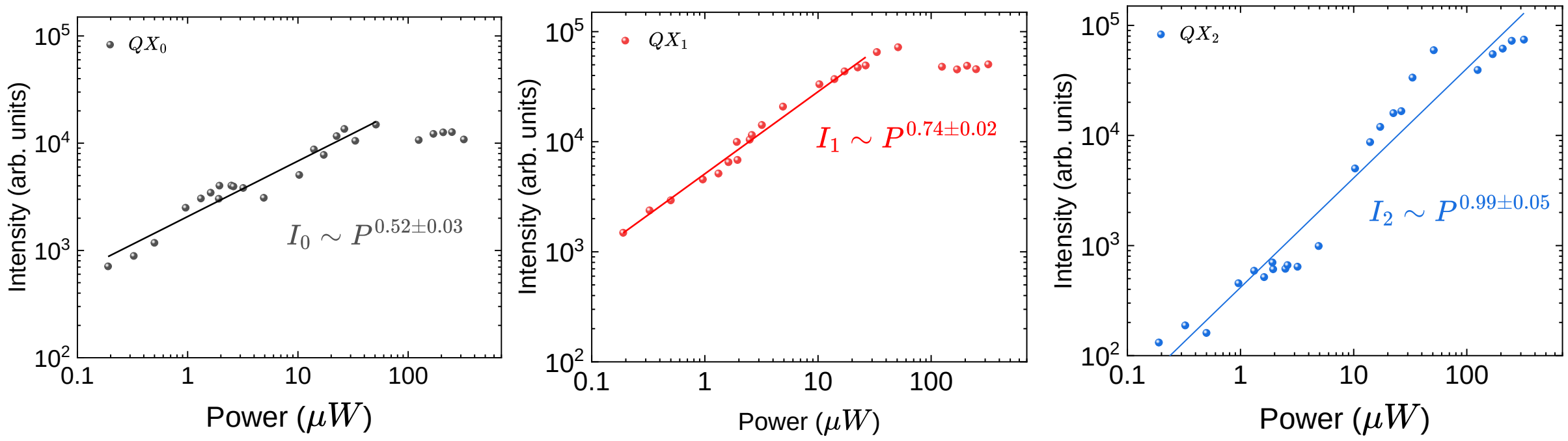


Figure SF9: Power dependent intensity of different quadrupolar States at F = 0, from lowest energy ($QX_0$, left) to highest energy ($QX_2$, right). The power law exponent is extracted by linear fitting the log of the intensity with respect to the log of the incident power in the non-saturated region. The weaker power dependence for lower energy states suggests longer lifetime and lower degeneracies for lower energy states. This enhanced lifetime for lower energy states likely result from suppressed nonradiative pathways and higher degree of localisation.

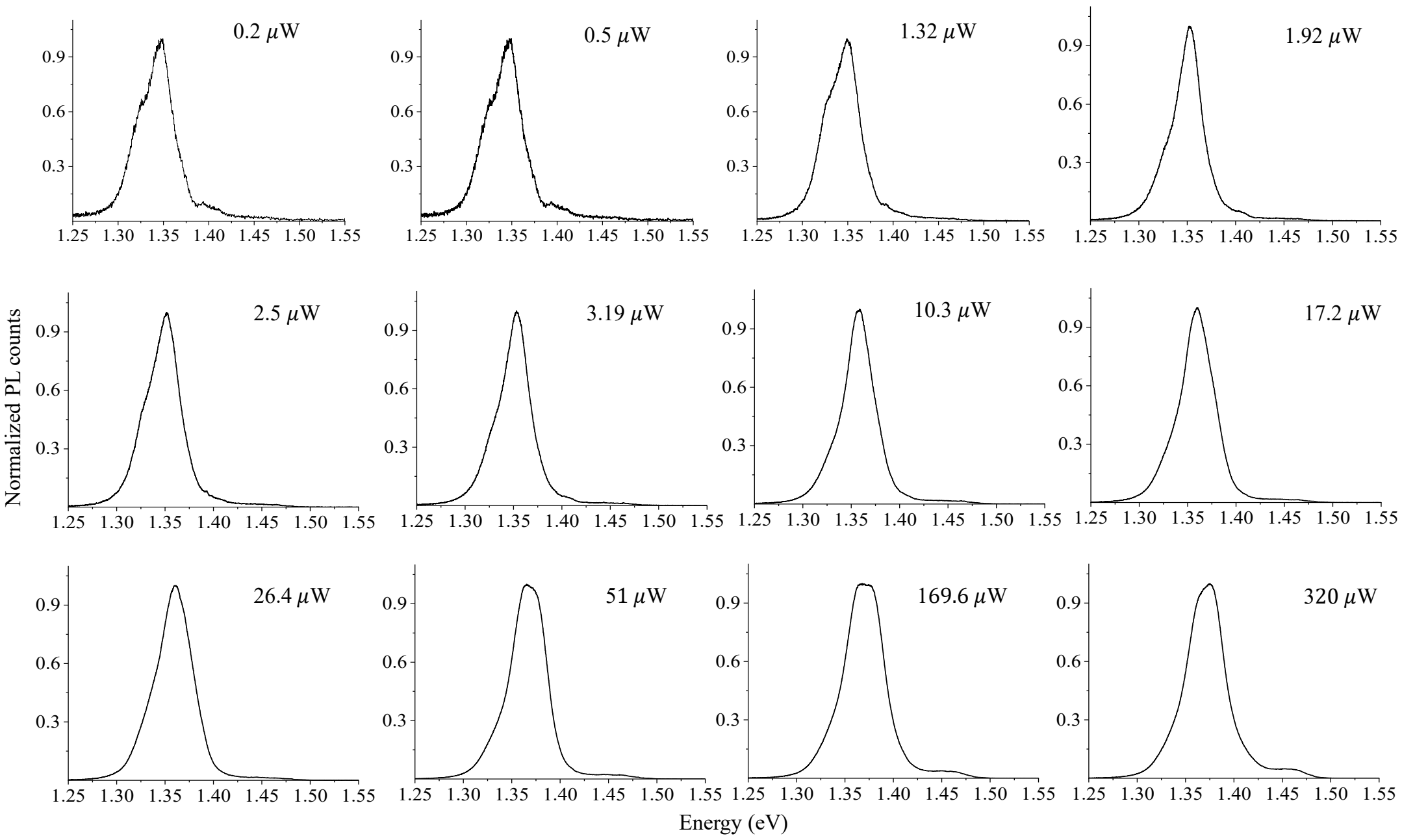


Figure SF10: PL spectra at each individual power value at F = 0. The excitation power values are noted at the right corner in each plot.

## S9 Finer resolution trilayer electric field dependence

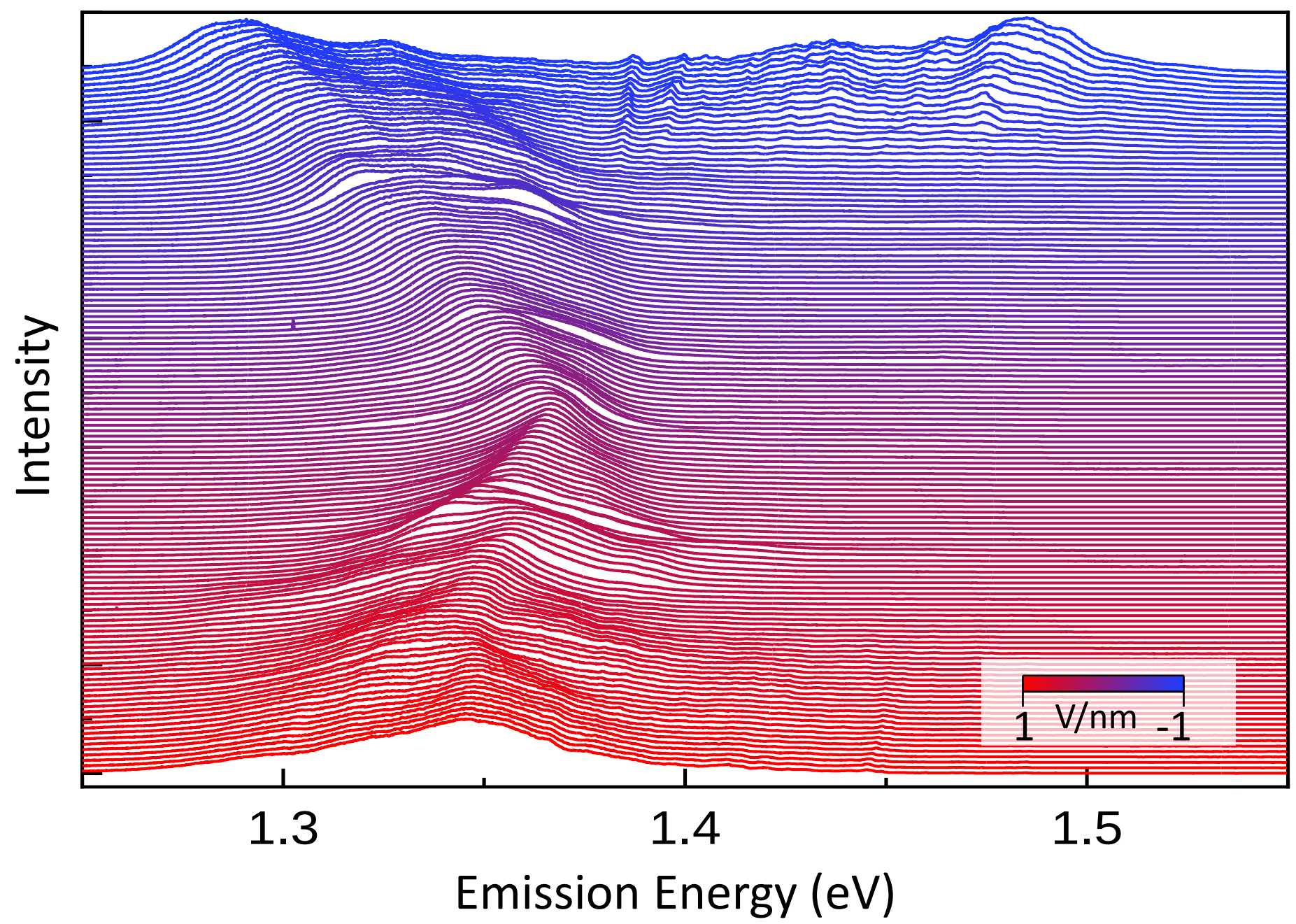


Figure SF11: Stacked plot of electric field dependent emission spectra from trilayer with finer electric field changes.

## S10 Effect of coupling constant on the energy-field slope (dipole moment) at high field

We estimate here the relative slopes of different moiré branches. For simplicity of the analysis and to gain an analytical solution, we assume that interaction between $IX_{ui}$ and $IX_{dj} (i \neq j)$ is negligible. The energy of the bright (symmetric) state is $\mathcal{E} \sim -\sqrt{(e.d.F)^2 + t^2}$. The effective dipole moment is

$$|\frac{\partial \mathcal{E}}{\partial F}| = \frac{ed}{\sqrt{1 + \left(\frac{t}{eFd}\right)^2}}$$

At higher field values $\frac{t}{eFd} << 1$, thus,

$$|\frac{\partial \mathcal{E}}{\partial F}| \approx ed \left(1 - \frac{1}{2}\left(\frac{t}{eFd}\right)^2\right)$$

For a given value of $F$, states with larger coupling constant will show smaller dipole moments and vice versa. We now note that the energy barrier between the moiré bands of the two isolated moiré systems ($IX_{ui}$ and $IX_{di}$) will be lower for the higher energy bands. Recall that $t_i$ is the coupling parameter between $IX_{ui}$ and $IX_{di}$. Since the coupling constant is expected to be higher with a lower energy barrier, $t_0 < t_1 < t_2$. Thus,

$$\left(\frac{\partial \mathcal{E}}{\partial F}\right)_{QX_0} > \left(\frac{\partial \mathcal{E}}{\partial F}\right)_{QX_1} > \left(\frac{\partial \mathcal{E}}{\partial F}\right)_{QX_2}$$

## S11 Conduction band hybridization and field dependent overlap of electron and hole wavefunction

The electrons and holes are modeled as particles confined in a 2D potential well. The shape of the potential well follows from the Type-II band alignment in $WS_2$ /$WSe_2$ /$WS_2$ heterostructure, and the band-tilting effect of the electric field is included. The electron wavefunction is solved numerically for each dipolar state for the field-dependent potential well at each field value. Let $|\psi_e^{u/d}(F)\rangle$ be the electron wavefunction for $u$ and $d$ dipolar excitons shown in the figure.

Next, we model the coupling between the two conduction bands, leading to the hybridization of the two dipolar excitons through the following Hamiltonian in the basis of $|\psi_e^{u/d}(F)\rangle$,

$$H = \begin{pmatrix} -e.d.F & -t_i(F) \\ -t_i(F) & e.d.F \end{pmatrix}$$

where $t_i(F)$ is the conduction band coupling parameter, which in general depends on the electric field $F$ and the band index $i$.

The eigenvalues of the above Hamiltonian, $E_{\pm} = \mp\sqrt{(e.d.F)^2 + t_i^2(F)}$ give the change in energies of the resulting hybridized states compared to the non-hybridized degenerate conduction bands. We refer to the state with energy change $E_+$ and $E_-$ as symmetric ($|\psi_e^S(F)\rangle$) and asymmetric states ($|\psi_e^A(F)\rangle$), respectively. Next, we find the eigenvectors of the Hamiltonian to reconstruct the hybridized wavefunctions as a superposition of $|\psi_e^u(F)\rangle$ and $|\psi_e^d(F)\rangle$. The (non-normalised) eigenvectors for the symmetric ($\lambda_S$) and asymmetric ($\lambda_A$) states are calculated to be

$$\lambda_S = \begin{bmatrix} \lambda_S^u \\ 1 \end{bmatrix}, \lambda_A = \begin{bmatrix} \lambda_A^u \\ 1 \end{bmatrix}$$

where $\lambda_S^u = \frac{e.d.F+\sqrt{(e.d.F)^2+t^2}}{t}$ and $\lambda_A^u = \frac{e.d.F-\sqrt{(e.d.F)^2+t^2}}{t}$. The (normalised) hybridized

wavefunction can now be stated as,

$$|\psi_e^S(F)\rangle = \frac{\lambda_S^u\,|\psi_e^u(F)\rangle + |\psi_e^d(F)\rangle}{\sqrt{\langle\psi_e^S(F)|\psi_e^S(F)\rangle}}, |\psi_e^A(F)\rangle = \frac{\lambda_A^u\,|\psi_e^u(F)\rangle + |\psi_e^d(F)\rangle}{\sqrt{\langle\psi_e^A(F)|\psi_e^A(F)\rangle}}$$

We calculate the field-dependent hole wavefunction in the trilayer system as $|\psi_h(F)\rangle$. The overlap between electron and hole wavefunction of the resulting hybridized states is given as,

$$\mathcal{O}_{S/A}(F) = \left|\langle\psi_e^{S/A}(F)|\psi_h(F)\rangle\right|^2$$

## S12 Anti-symmetric branch at high electric field showing multiple peaks

A zoomed-in spectra at the high field region is reproduced in **Supplementary Fig. SF12**, showing emergence of multiple peaks. As discussed in the main text, the emission area increases at high electric field due to lack of exciton funneling to supermoiré points. The increased emission area makes the emission more sensitive to sample inhomogeneities. An inhomogeneous moiré landscape can generate a spatially varying in-plane electric field from the applied vertical field.[6] This in turn locally breaks the rotational symmetry of the 2D harmonic moiré wells, thus lifting the degeneracy of the higher energy states. This explains the emergence of multiple peaks at high field in the higher energy resonances.

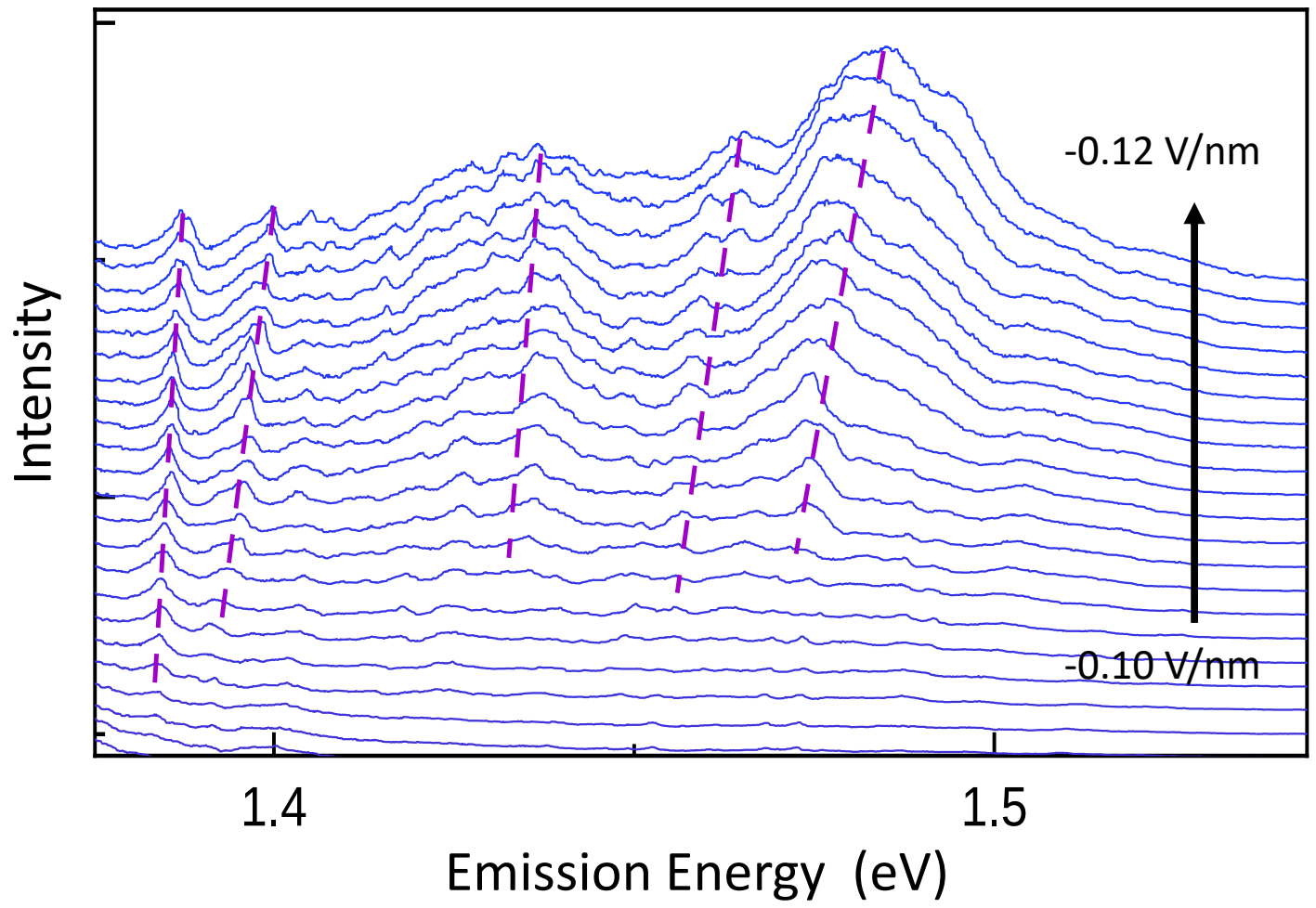


Figure SF12: External electric field dependent PL spectra for the higher energy (anti-symmetric) quadrupolar branch at high electric field values. The dashed lines are a guide to the eye to show the blue shift with electric field.

## S13 Power dependence at different electric fields

Power dependence of PL emission from trilayer is shown in **Supplementary Fig. SF13** for different electric field values. Blue-shift in emission with power signifies exciton-exciton interaction with increasing excitonic density. At zero electric field (**Supplementary Fig. SF13 (b,e)**), the emission comes from quadrupolar excitons. At higher excitonic density the ground state configuration is that of staggered phase due to its attractive nature. With large negative field (**Supplementary Fig. SF13 (c,f)**) the trilayer become an effective bilayer with most of the electronic wavefunction residing in only one of the $WS_2$ layers. Major weight of the electronic wavefunction resides in the top layer for symmetric and in the bottom layer anti-symmetric state, respectively. Similar magnitude of blueshift in both low and high energy emission hints towards supermoiré confinement. Assuming purely coulomb interaction between the excitons, the ratio of dipolar excitonic interaction to quadrupolar interaction ($U_{bi}$ / $U_{tri}$) is density dependent.[2] This ratio can range from 1.5 and up as shown in the **Supplementary Fig. SF13 (g)**. This higher shift in dipolar exciton with respect to the quadrupolar exciton is assuming that the separation between dipolar excitons and that between the quadrupolar excitons is the same. This does not hold when the localization length scale is different for the two, given the $1/r^2$ interaction. Due to funnelling of quadrupolar excitons funnelling into supermoiré, inter-particle distance between the quadrupolar excitons will be smaller. This agrees well with the observation of smaller difference between the dipolar and quadrupolar excitons for similar range of excitation power. On the other hand, at high electric field, in the absence of any such funneling, the entire sample emits, reducing the relative exciton density.

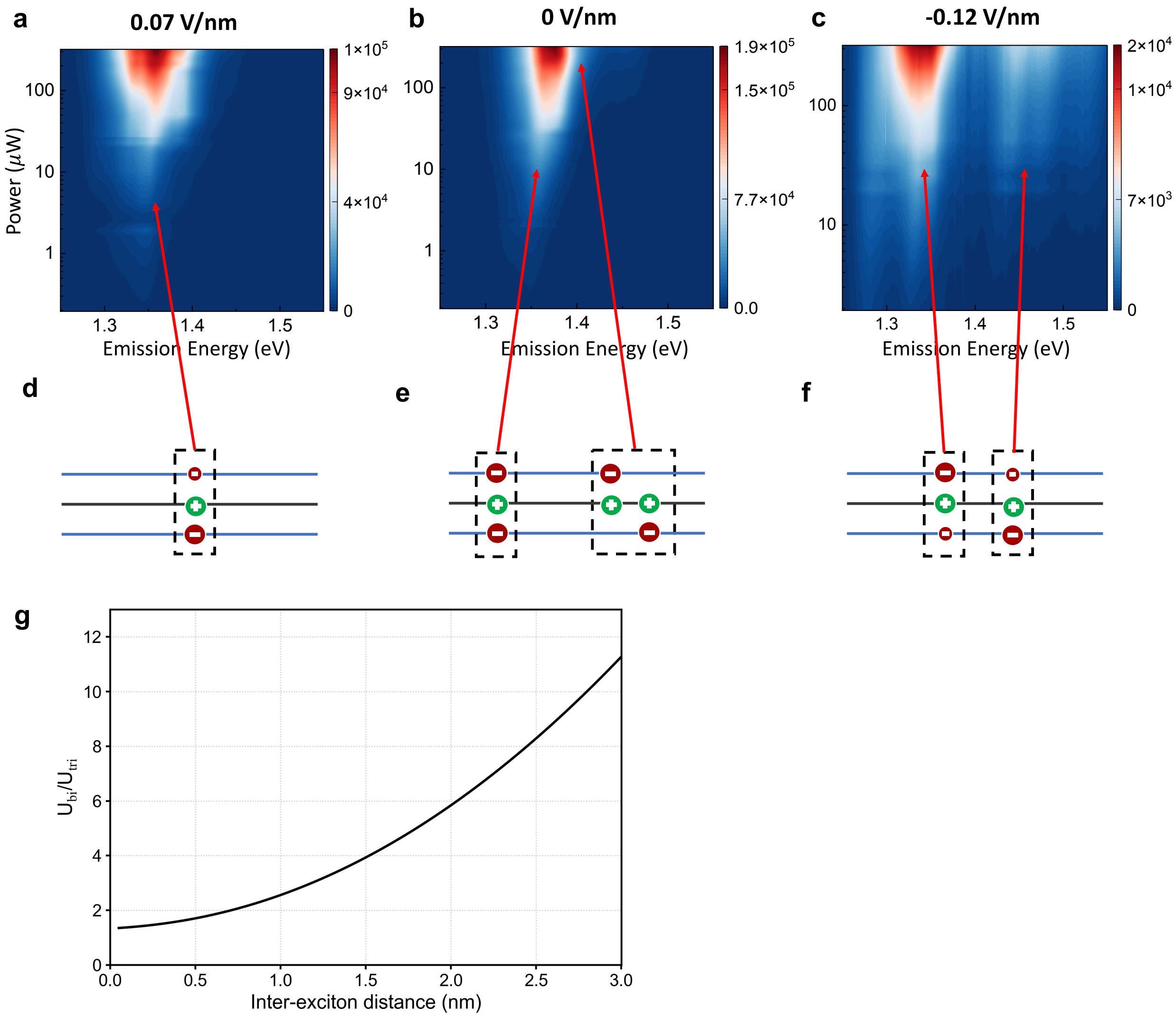


Figure SF13: Power dependent PL spectra at external electric field of **(a)** 0.07 V/nm **(b)** 0 V/nm **(c)** -0.12 V/nm. **(d,e,f)** Illustration for electron-hole configuration for different electric field and excitation powers. The arrows map the electron-hole configuration inside the dashed-box to the corresponding electric field and excitation power range.**(g)** Ratio of coulomb interaction potential between dipolar and quadrupolar excitons as a function of inter-exciton distance.

## S14 Polarization resolved trilayer emission

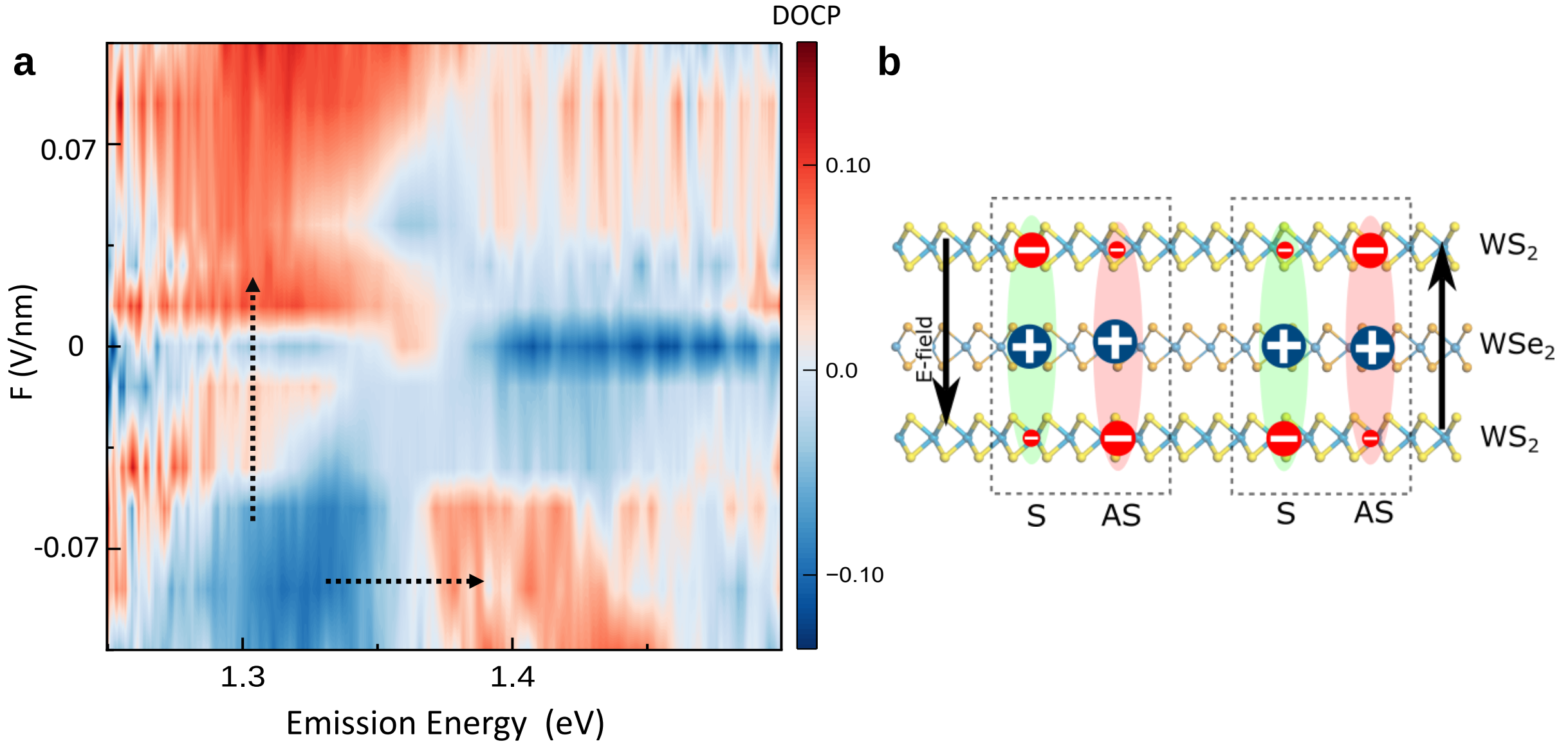


Figure SF14: **Polarization with electric field:** **(a)** Variation of the degree of circular polarization (DOCP=$\frac{I_{co}-I_{cross}}{I_{co}+I_{cross}}$) of the emission from trilayer as a function of emission energy and applied electric field. The dashed arrows show the direction of polarization flipping with an electric field (vertical arrow) and emission energy (horizontal arrow). **(b)** The schematic shows the distribution of the electrons (red circles) and holes (blue circles) in positive and negative field direction for the S (symmetric $QX_S$) and AS (anti-symmetric $QX_A$) states. The black arrow adjacent to the dashed box shows the direction of the applied electric field for states represented inside the box.

We perform circular polarization-resolved PL measurements. We plot the degree of circular polarization (DOCP) of the emission as a function of electric field and emission energy in a color plot in **Fig. SF14a**. For the $QX_S$ branch, a crossover from co- to cross-polarized emission is observed with a change in the direction of the electric field. Moreover, at a given electric field, the DOCP switches from negative to positive between the low and high energy branches. These observations suggest that the high energy emission at negative field and low energy emission at positive field originate from the same (top) moiré layer. This is in agreement with the assertion that the lower and higher energy peaks originate from the $QX_S$ and $QX_A$ branches of the quadrupolar exciton, as schematically explained in **Fig. SF14b**.


\end{document}